\documentclass[fleqn,usenatbib]{mnras}

\usepackage{newtxtext,newtxmath}
\usepackage{color,soul}

\usepackage[T1]{fontenc}

\DeclareRobustCommand{\VAN}[3]{#2}
\let\VANthebibliography\thebibliography
\def\thebibliography{\DeclareRobustCommand{\VAN}[3]{##3}\VANthebibliography}

\usepackage{graphicx}	
\usepackage{amsmath}	
\usepackage{float}

\newcommand*{\rom}[1]{\expandafter\romannumeral #1}

\newcommand\chem[2]{#1$\;${%
\ifx\@currsize\normalsize\small \else
\ifx\@currsize\small\footnotesize \else
\ifx\@currsize\footnotesize\scriptsize \else
\ifx\@currsize\scriptsize\tiny \else
\ifx\@currsize\large\normalsize \else
\ifx\@currsize\Large\large
\fi\fi\fi\fi\fi\fi
\rom{#2}}\relax}%

\newcommand{\oiii}{\lbrack $\rm O \, {\sc III}$\rbrack}
\newcommand{\cii}{[C\ensuremath{\,\textsc{ii}}]}

\title[The host dark matter haloes of the first quasars]{The host dark matter haloes of the first quasars} 

\author[Costa, Tiago]{
Tiago Costa\thanks{E-mail: tiago.costa@newcastle.ac.uk}
\\
Max-Planck-Institute for Astrophysics, Karl-Schwarzschildstra{\ss}e 1, 85748 Garching bei M\"unchen, Germany \\
School of Mathematics, Statistics and Physics, Newcastle University, Newcastle upon Tyne, NE1 7RU, UK
}

\date{Accepted XXX. Received YYY; in original form ZZZ}

\pubyear{2015}

\begin{document}
\label{firstpage}
\pagerange{\pageref{firstpage}--\pageref{lastpage}}
\maketitle

\begin{abstract}
If $z > 6$ quasars reside in rare, massive haloes, $\Lambda$CDM cosmology predicts they should be surrounded by an anomalously high number of bright companion galaxies. Here I show that these companion galaxies should also move unusually fast. Using a new suite of cosmological, `zoom-in' hydrodynamic simulations, I present predictions for the velocity distribution of quasar companion galaxies and its variation with quasar host halo mass at $z \, = \, 6$. Satellites accelerate as they approach the quasar host galaxy, producing a line-of-sight velocity profile that broadens with decreasing distance to the quasar host galaxy. This increase in velocity dispersion is particularly pronounced if the host halo mass is $\gtrsim 5 \times 10^{12} \, \rm M_\odot$. Typical line-of-sight speeds rise to $\approx 500 \, \rm km \, s^{-1}$ at projected radii $\sim 10 \, \rm kpc$. For about $10\%$ of satellites, they should exceed $800 \, \rm km \, s^{-1}$, with $\approx 5\%$ of companions reaching line-of-sight speeds $\sim 1000 \, \rm km \, s^{-1}$. For lower host halo masses $\approx 5 \times 10^{11} \-- 10^{12} \, \rm M_\odot$, the velocity profile of companion galaxies is significantly flatter. In this case, typical line-of-sight velocities are $\approx 250 \, \rm km \, s^{-1}$ and do not exceed $\approx 500 \, \rm km \, s^{-1}$. A comparison with existing ALMA, JWST and MUSE line-of-sight velocity measurements reveals that observed $z > 6$ quasar companions closely follow the velocity distribution expected for a host halo with mass $\gtrsim 5 \times 10^{12} \, \rm M_\odot$, ruling out a light host halo. Finally, through an estimate of UV and \oiii \, luminosity functions, I show that the velocity distribution more reliably discriminates between halo mass than companion number counts, which are strongly affected by cosmic variance.
\end{abstract}




\section{Introduction}
Quasars have been detected out to $z \, \approx \, 7.6$ \citep{Mortlock:11, Banados:18, Yang:20, Wang:21}. The supermassive black holes powering $z > 6$ quasars have less than $\approx 1 \, \rm Gyr$ to grow to their estimated masses $\gtrsim 10^9 \, \rm M_\odot$.
Even if they grow from massive \emph{seed} black holes, accretion flows must ensure these supermassive black holes either accrete at their Eddington limit almost constantly \citep{Sijacki:09, Costa:14} or undergo periods of sustained super-Eddington accretion \citep{Bennett:23, Schneider:23}. 
Within the $\Lambda$CDM framework, either scenario requires that early supermassive black holes form and grow in the centres of rare, massive dark matter haloes \citep{Efstathiou:88, Volonteri:06, Costa:14}. 
The required dark matter halo mass $M_{\rm h}$ can be estimated as
\begin{equation}
\begin{aligned}
    M_{\rm h} \, \approx \, 1.3 \times 10^{13} \, \left( \frac{M_{\rm BH}}{10^9 \, \rm M_\odot} \right) & \left( \frac{M_{\rm BH} / M_\star}{0.005} \right)^{-1} \\
    & \times \left( \frac{M_\star / \left(f_{\rm b} M_{\rm h}\right)}{0.1} \right)^{-1} \left( \frac{f_{\rm b}}{0.16} \right)^{-1}  \rm M_\odot \, ,
\end{aligned}
\label{eq:halomass}
\end{equation}
where $M_{\rm BH}$, $M_\star$ and $f_{\rm b}$ denote black hole mass, stellar mass and cosmic baryon fraction, respectively. The black hole -- stellar mass ratio appearing in Equation~\ref{eq:halomass} is normalised to the local value \citep{Kormendy:13}, while the stellar -- baryonic mass ratio is set to a value approximately in line with abundance matching predictions \citep[e.g.][]{Behroozi:18}. While abundance matching constraints are highly uncertain at $z \, = \, 6$, Equation~\ref{eq:halomass} shows that lowering the quasar host halo mass requires invoking higher-than-expected stellar -- baryonic mass ratios or for the black hole -- stellar mass ratio to exceed the local relation. 

The theoretical prediction that $z > 6$ quasars reside in massive haloes, however, still requires observational validation. 
An indirect probe of the host halo mass is its associated matter overdensity.
At $z > 6$, haloes with mass $\gtrsim 10^{12} \, \rm M_\odot$ are rare and must form in high $\sigma$-peaks of the cosmological density field.
Consequently, massive galaxies hosting $z > 6$ quasars should contain in their surroundings an anomalously high number of companion galaxies \citep{Munoz:08}. 
The search for galaxies in quasar fields at $z > 6$ has, however, not led to a conclusive picture. Companion detection techniques involving, for instance, photometric selection of Lyman Break galaxies \citep{Willott:05, Zheng:06, Simpson:14, Morselli:14, Mignoli:20, Champagne:23} or sub-mm galaxies \citep{Li:23} in deep pencil-beam surveys and the detection of Ly$\alpha$ emitters in narrow-band observations \citep{Banados:13, Farina:17, Mazzucchelli:17} reveal quasar environments ranging from overdense \citep[e.g.][]{Balmaverde:17, Meyer:22, Overzier:22} to underdense \citep[e.g.][]{Kim:09, McGreer:14, Ota:18}. Recent observations with the James Webb Space Telescope (JWST) find clear enhancements of \oiii \, emitters around $z > 6$ quasars \citep{Wang:23, Kashino:23}, but only two fields have been examined so far. 

The question of why galaxy overdensities around $z > 6$ quasars are not detected more often continues to generate debate. Various explanations have been put forward, including
\begin{enumerate}
    \item the possibility that many, perhaps most, quasar companions are dust-obscured \citep[e.g.][]{Mazzucchelli:17},
    \item the possibility that quasars suppress galaxy formation in their vicinity \citep{Utsumi:10}. The few theoretical studies that have explored such a scenario, however, either find that the negative impact of active galactic nucleus (AGN) feedback is restricted to the very faintest satellites \citep{Costa:14, Chen:20}, or even that this results in a brighter companion population \citep{Zana:22},
    \item the host haloes of $z > 6$ quasars are light, with virial masses $\lesssim 10^{12} \, \rm M_\odot$ \citep[e.g.][]{Fanidakis:13}, defying expectations.
\end{enumerate}

Another possible explanation is cosmic variance. The magnitude of an overdensity depends on the spatial scale in which it is defined.
\citet{Angulo:12} show that while haloes with masses $(1 \-- 10) \times 10^{12} \, \rm M_\odot$ at $z > 6$ typically trace overdensities at scales $\lesssim 10 \, \rm Mpc$ (comoving), these may be embedded in either underdensities or overdensities on larger scales, potentially explaining the ambiguous observational picture. 

This paper explores a different, more direct probe of the gravitational potential wells hosting $z > 6$ quasars.
My reasoning follows findings presented in \citet{Costa:15}, where it is shown that the steep gravitational potentials of $z > 6$ massive haloes accelerate infalling gas and satellites to extreme speeds of $\sim 1000 \, \rm km \, s^{-1}$. This finding hints at an alternative observational signature of a rare, massive dark matter halo at $z > 6$: one should look for satellites in the vicinity of $z > 6$ quasars with unusually high line-of-sight velocities. Such an approach resembles the long-known strategy of constraining galactic halo masses via satellite dynamics \citep{Little:87, Prada:03, vandenBosch:04, Tempel:06, Conroy:07}.
In this paper, this idea is applied in the context of $z > 6$ quasar host haloes. Length scales are given in physical coordinates. Magnitudes are quoted in the AB system.

\section{Numerical Simulations}
\label{sec:simulations}
The six most massive dark matter haloes of the Millennium volume \citep{Springel:05} are targeted at $z \, = \, 6$. These haloes have virial masses\footnote{Virial masses and radii are defined within a spherical region enclosing a mean density 200 times the critical density.} of $\left( 6 \-- 7 \right) \times 10^{12} \, \rm M_\odot$ (see Table~\ref{tab:number_counts}) and have been singled out in previous studies \citep{Li:07, Sijacki:09,Costa:14,Zhu:22,Bennett:23} as the likely hosts of $\sim 10^9 \, \rm M_\odot$ black holes at $z \, = \, 6$. Virial radii range from $\approx 83 \, \rm kpc$ to $\approx 87 \, \rm kpc$. 

In order to quantify the halo mass dependence of the satellite velocity distribution, the evolution of six haloes with masses in the range $M_{\rm vir} \, = \, \left( 5 \-- 10 \right) \times 10^{11} \, \rm M_\odot$ is followed as well. While these haloes also trace overdensities, they are expected to host lower mass black holes $\lesssim 10^8 \, \rm M_\odot$ according to cosmological simulations \citep[see][]{Costa:14}. In this paper, these two samples are respectively referred to as `high-mass' (\texttt{HM}) or `low-mass' (\texttt{LM}) host halo samples. When referring to individual haloes from either sample, I use the nomenclature \texttt{LM-3} e.g. to refer to halo 3 of the `low-mass' sample or \texttt{HM-6} for halo 6 of the `high-mass' sample. 

Note that in hydrodynamic simulations, the black hole mass reached in a given dark matter halo \textit{by a certain redshift} is sensitive to a variety of numerical parameters, including the maximum allowed Eddington ratio and AGN feedback efficiency \citep[see][for examples of how parameter variations result in order of magnitude variations in black hole mass]{Zhu:22, Bennett:23}. Therefore each halo is here treated as a potential host of a bright $z > 6$ quasar. The question is then asked: \textit{what is the velocity distribution predicted for the two different halo samples and how does this compare with observed data?}

\begin{table}
	\centering
	\caption{List of simulations and number counts of galaxy companions within a radial distance of $2 \, \rm Mpc$ from the most massive black hole. The first column lists the simulations, the second quotes the virial mass of the target halo (evaluated within the radial scale at which the mean enclosed density is 200 times the critical density), while the third and fourth columns give number counts of companions with stellar masses $> 10^{10} \, \rm M_\odot$ and $> 10^{9} \, \rm M_\odot$, respectively. All but one of the simulated quasar fields possesses one or more massive galaxies with $M_\star > 10^{10} \, \rm M_\odot$ (excluding the quasar host galaxy itself).}
	\label{tab:number_counts}
	\begin{tabular}{lrcc} 
		\hline
		Sim. & $M_{\rm vir} \, \rm [M_\odot]$ & $N \, (M_\star > 10^{10} \, \rm M_\odot)$ & $N \, (M_\star > 10^{9} \, \rm M_\odot)$\\
  		\hline
		\texttt{LM-1} & $6.4 \times 10^{11}$ & 0 & 4 \\
  		\texttt{LM-2} & $6.6 \times 10^{11}$ & 0 & 7 \\
		\texttt{LM-3} & $1.19 \times 10^{12}$ & 1 & 25\\
		\texttt{LM-4} & $9.6 \times 10^{11}$ & 0 & 12\\
		\texttt{LM-5} & $9.8 \times 10^{11}$ & 0 & 24\\
		\texttt{LM-6} & $1.01 \times 10^{12}$ & 0 & 23\\
		\hline
		\texttt{HM-1} & $6.75 \times 10^{12}$ & 2 & 31\\
  		\texttt{HM-2} & $6.19 \times 10^{12}$ & 0 & 45\\
		\texttt{HM-3} & $6.22 \times 10^{12}$ & 5 & 31\\
		\texttt{HM-4} & $6.21 \times 10^{12}$ & 2 & 20\\
		\texttt{HM-5} & $7.02 \times 10^{12}$ & 1 & 16\\
		\texttt{HM-6} & $6.96 \times 10^{12}$ & 10 & 116\\
		\hline
	\end{tabular}
\end{table}

To retain their predictive character, the simulations presented here adopt a very close numerical setup to that presented in \citet{Costa:14} and \citet{Costa:15}. They are `zoom-in' simulations: high-resolution is placed in a Lagrangian patch traced back to the initial conditions, while the remainder of the Millennium volume is simulated at lower resolution. At $z \, = \, 6$, the approximately spherical high-resolution region have radii $\approx 2\, \rm Mpc$ for the \texttt{HM} haloes and $\approx 3 \, \rm Mpc$ for \texttt{LM} haloes.

The simulations are performed with the moving-mesh code \textsc{Arepo} \citep{Springel:10, Pakmor:16, Weinberger:20}. \textsc{Arepo} solves the Euler equations of hydrodynamics on an irregular mesh constructed through a Voronoi tessellation of a number of mesh-generating points. The flow is linearly reconstructed within each Voronoi cell, giving second order spatial accuracy. Additionally, mesh-generating points are advected with the flow, endowing the solver with Lagrangian properties. N-body dynamics of dark matter, stellar populations, gas and black holes is computed via a TreePM method \citep{Springel:05}.

\begin{table*}
	\centering
	\caption{Summary of quasar satellite galaxy sample, showing quasar ID, redshift, apparent magnitude in the J band, estimated black hole mass, the number of detected companion galaxies (with $|\Delta v| < 4000 \, \rm km \, s^{-1}$) broken down according to their detection method and the publications from which the companion properties, central black hole mass and UV ($1450$ \AA) magnitude were obtained. There are 79 (65) galaxies with $|\Delta v| < 4000 \, \rm km \, s^{-1}$ ($|\Delta v| < 1300 \, \rm km \, s^{-1}$) in 18 quasars at $z > 6$. The population of quasars examined here is in line with brightest quasars at $z \, > \, 6$ and most are thought to be powered by black holes with mass $> 10^9 \, \rm M_\odot$.}
	\label{tab:galaxySample}
	\begin{tabular}{lccccp{0.43\linewidth}} 
		\hline
		QSO ID & $z_{\rm QSO}$ & $M_{\rm UV}$ & $M_{\rm BH} \, [M_\odot]$ & \# companions & References \\
  		\hline
        J0020+3653 & $6.834$ & $-26.92$ & $2.90 \times 10^9$ & [OIII]: 3 & sats. and $M_{\rm BH}$ \citep{Marshall:23}, $M_{\rm UV}$ \citep{Farina:22} \\
        J0100+2802 & $6.327$ & $-29.02$ & $1.584 \times 10^{10}$ & [CII]: 1, [OIII]: 31 & sats. \citep{Venemans:20, Kashino:23}, $M_{\rm BH}$ \citep{Eilers:23}, $M_{\rm UV}$ \citep{Farina:22} \\
        J0411+0907 & $6.824$ & $-26.58$ & $1.85 \times 10^9$ & [OIII]: 1 & sats. and $M_{\rm BH}$ \citep{Marshall:23}, $M_{\rm UV}$ \citep{Farina:22} \\
  		J0842+1218 & $6.076$ & $-26.69$ & $2.54 \times 10^9$ & [CII]: 2 & sats. \citep{Decarli:17}, $M_{\rm BH}$ and $M_{\rm UV}$ \citep{Farina:22}  \\
		J0305+3150 & $6.614$ & $-25.91$ & $6.200 \times 10^8$ & [CII]: 3, Ly$\alpha$: 1, [OIII]: 17 & sats. \citep{Farina:17, Venemans:19, Wang:23}, $M_{\rm BH}$ \citep{Yang:23}, $m_{\rm J}$ \citep{Farina:22} \\
        J1030+0524 & $6.308$ & $-26.76$ & $1.93 \times 10^9$ & Ly$\alpha$: 2 & sats. \citep{Mignoli:20}, $M_{\rm BH}$ and $M_{\rm UV}$ \citep{Farina:22} \\
    	J1306+0356 & $6.033$ & $-26.70$ & $1.95 \times 10^9$ & [CII]: 1 & sats. \citep{Neeleman:19}, $M_{\rm BH}$ \citep{Mazzucchelli:23}, $M_{\rm UV}$ \citep{Farina:22} \\
		J1319+0950 & $6.133$ & $-26.80$ & $3.39 \times 10^9$ & [CII]: 2 & sats. \citep{Venemans:20, Neeleman:21}, $M_{\rm BH}$ \citep{Mazzucchelli:23}, $M_{\rm UV}$ \citep{Farina:22} \\
        J1342+0928 & $7.534$ & $-26.64$ & $1.97 \times 10^9$ & [CII]: 1 & sats. \citep{Venemans:20}, $M_{\rm BH}$ and $M_{\rm UV}$ \citep{Farina:22} \\
        J2054+0005 & $6.038$ & $-26.15$ & $1.48 \times 10^9$ & [CII]: 1 & sats. \citep{Venemans:20}, $M_{\rm BH}$ and $M_{\rm UV}$ \citep{Farina:22} \\
  		J2100+1715 & $6.081$ & $-24.63$ & $4.49 \times 10^9$ & [CII]: 1 & sats. \citep{Decarli:17}, $M_{\rm BH}$ and $M_{\rm UV}$ \citep{Farina:22} \\
		J2318+3113 & $6.444$ & $-26.11$ & $1.45 \times 10^9$ & [CII]: 4 & sats. \citep{Venemans:20, Neeleman:21}, $M_{\rm BH}$ and $M_{\rm UV}$ \citep{Farina:22} \\
  		P009+10   & $6.004$ & $-26.03$ & $2.35 \times 10^9$ & [CII]: 1 & sats. \citep{Neeleman:21}, $M_{\rm BH}$ and $M_{\rm UV}$ \citep{Farina:22}  \\
    	P065+26   & $6.189$ & $-26.94$ & $3.63 \times 10^9$ & [CII]: 1 & sats. \citep{Neeleman:21}, , $M_{\rm BH}$ \citep{Mazzucchelli:23}, $M_{\rm UV}$ \citep{Farina:22} \\
    	P167+13   & $6.515$ & $-25.58$ & $4.900 \times 10^8 $ & [CII]: 1 & sats. \citep{Neeleman:19}, $M_{\rm BH}$ and $m_{\rm J}$ \citep{Venemans:15} \\
        P183+05   & $6.439$ & $-26.87$ & $2.57 \times 10^9$ & [CII]: 1 & sats. \citep{Venemans:20}, $M_{\rm BH}$ \citep{Mazzucchelli:23}, $M_{\rm UV}$ \citep{Farina:22} \\
     	P231+20   & $6.587$ & $-27.07$ & $3.31 \times 10^9$ & [CII]: 2, Ly$\alpha$: 2 & sats. \citep{Decarli:17, Neeleman:19, Meyer:22}, $M_{\rm BH}$ \citep{Mazzucchelli:23}, $M_{\rm UV}$ \citep{Farina:22} \\
    	P308+21   & $6.234$ & $-26.27$ & $1.23 \times 10^9$ & [CII]: 2 & sats. \citep{Decarli:17}, $M_{\rm BH}$ \citep{Mazzucchelli:23}, $M_{\rm UV}$ \citep{Farina:22} \\
		\hline
	\end{tabular}
\end{table*}

The simulations follow primordial and metal-line cooling down to $T \, = \, 10^4 \, \rm K$, assuming photo-ionisation equilibrium against the spatially homogeneous UV background of \citet{Faucher-Giguere:10}. Star formation is modelled following \citet{Springel:03}. As in \citet{Costa:14}, `heavy' black hole seeds with mass $\approx 10^5 \, \rm M_\odot$, as expected from `direct collapse' scenarios \citep[e.g.][]{Inayoshi:20}, are placed into the centres of haloes once they exceed a mass $10^{10} \, \rm M_\odot$. Black hole growth then proceeds via black hole mergers and, mainly, via gas accretion, modelled as a Bondi-Hoyle flow with an accretion rate capped at the Eddington limit. Of central importance to the satellite population is feedback from supernovae, here modelled following \citet{Springel:03}. \citet{Costa:14} show this can sensitively regulate the stellar mass- and UV luminosity functions at $z > 6$. Here, mass is ejected from star-forming regions at a rate equal to the star formation rate, i.e. a mass-loading factor $\eta \, = \, 1$ is adopted, with an initial speed $v_{\rm w} \approx 1700 \, \rm km \, ^{-1}$. As it sweeps-up ambient gas, the wind slows down, so choosing a high wind speed is no guarantee it will escape the halo. 
Note also that while the feedback strength affects the number count of observable satellites, it does not affect the bulk velocity of satellite galaxies, which is driven by gravitational interactions with the host halo. In Section~\ref{sec:luminosity_functions}, I show that this model reproduces the UV luminosity function observed at $z \, = \, 6$ for the two least-biased `zoom-in' regions of the `low-mass' sample (where closest agreement may be expected).

One difference with respect to \citet{Costa:15} is the improved numerical resolution of the new simulations, which yield a better-resolved satellite population.
The mass resolution of dark matter particles is improved by a factor 8, such that $m_{\rm dm} \, \approx \, 1.4 \times 10^6 \, \rm M_\odot$, as is the target mass of gas cells, which becomes $m_{\rm gas} \, \approx \, 2 \times 10^5 \, \rm M_\odot$. Haloes with $M_{\rm vir} \approx 5 \times 10^7 \, \rm M_\odot$ are thus resolved with 32 dark matter particles. These improvements justify decreasing the \textit{maximum} softening lengths by a factor two to $500 \, \rm pc$ (comoving). In {\sc Arepo}, cells advect with the flow and are refined/de-refined to ensure a roughly constant mass per cell. Consequently, there is a hierarchy of cell sizes. If size is defined as the radius of a sphere with a volume equal to the actual cell volume, this ranges from $\approx 5 \, \rm pc$ within the quasar host galaxy to $\approx 660 \, \rm pc$ in the hot, diffuse outskirts of the halo. About $10\%$ of cells within the central $10 \, \rm kpc$ have sizes $< 10 \, \rm pc$, with a minimum of  $\approx 3 \, \rm pc$.

Like the parent Millennium volume, I use cosmological parameters $\Omega_{\rm \Lambda} \, = \, 0.75$, $\Omega_{\rm m} \, = \, 0.25$, $\Omega_{\rm b} \, = \, 0.045$, $\sigma_{\rm 8} \, = \, 0.9$ and $h \, = \, 0.73$, in agreement with Wilkinson Microwave Anisotropy Probe (WMAP) constraints \citep{Spergel:07}.

\section{Observations}
\label{sec:observations}

The detection of quasar companions has been achieved through a variety of observations, including the detection of CO \citep{Wang:10}, rest-FIR continuum or \cii\, emission \citep[e.g.][]{Trakhtenbrot:17, Willott:17, Decarli:17, Bischetti:18, Izumi:21}, Ly$\alpha$ \citep[e.g.][]{Farina:17} and \oiii\, emission \citep{Wang:23, Kashino:23}. To compare the theoretical predictions (presented in Section~\ref{sec:results}) with observations, measurements of line-of-sight velocities and projected radial distances are gathered from the literature. 

\subsection{Galaxies in quasar environments}
\label{sec:dataset}
The sample of quasar companions built here includes data for galaxies detected as
\begin{enumerate}
    \item Ly$\alpha$ emitters found using the Multi Unit Spectroscopic Explorer (MUSE) in \citet{Farina:17}, \citet{Mignoli:20} and \citet{Meyer:22}, 
    \item \cii\, emitters from Atacama
Large Millimeter/submillimeter Array (ALMA) observations \citep{Decarli:17, Neeleman:19, Neeleman:21, Venemans:19, Venemans:20} and
    \item \oiii\, emitters from JWST \citep{Marshall:23}, including the ASPIRE \mbox{\citep{Wang:23}} and EIGER \mbox{\citep{Kashino:23}} projects. 
\end{enumerate}

The properties of the $z > 6$ quasars around which the companions in the sample have been detected is provided in \mbox{Table~\ref{tab:galaxySample}}, together with references to the publications presenting their discovery. 
The various instruments used in these various observations have different fields-of-view (FoV) and thus probe companions on different scales. The FoV is $1 \times 1 \, \rm arcmin^2$ for MUSE. For ALMA\footnote{ALMA Technical Handbook: \url{https://arc.iram.fr/documents/cycle7/ALMA_Cycle7_Technical_Handbook.pdf}}, the FoV is $\approx 1.13 \lambda / D$, where $\lambda$ is the observed-frame wavelength of the observation, assumed here to be that of the \cii \, emission line ($157.74 \, \rm \mu m$ in the rest-frame), and $D \, = \, 12 \, \rm m$ is the telescope diameter. In the redshift range considered here, the FoV of the ALMA observations is $\approx 25 \times 25 \, \rm arcsec^2$ \citep{Decarli:17}. The FoV is $3 \times 3 \, \rm arcsec^2$ for the JWST/NIRSpec integral field unit used in \citet{Marshall:23}. ASPIRE uses JWST/NIRCam, which has two modules, each covering $2.2 \times 2.2 \, \rm arcmin^2$, albeit at different sensitivity. \mbox{\citet{Wang:23}} estimate the effective survey area within their the higher-sensitivity module to be $5.5 \, \rm arcmin^2$. Finally, EIGER covers a $\approx 6 \times 3 \, \rm arcmin^2$ field \citep[e.g.][]{Eilers:23}.

In total, the sample contains 79 $z > 6$ quasar companions with $|v_{\rm los}| < 4000 \, \rm km \, s^{-1}$ and accurate spectroscopic redshifts with typical errors of $\Delta z \, \approx \, 10^{-4} \-- 5 \times 10^{-3}$.
These galaxies are found in the environments of 18 quasars. These quasars have absolute UV magnitudes (at $1450$ \AA) ranging from $-24.3$ to $\approx -29$ (see Table~\mbox{\ref{tab:galaxySample}}) and a mean of $-26.5$, in line with the brightest population of $z \, > \, 6$ quasars \citep[see e.g. Figure 1 in][]{Fan:23}. Note, for comparison, that the fainter $z \gtrsim 6$ quasar population discovered by Hyper Suprime-Cam typically have $M_{\rm UV} \gtrsim -24$ \citep{Matsuoka:22}. Estimated black hole mass estimates (Table~\mbox{\ref{tab:galaxySample}}) range from $5 \times 10^8 \, \rm M_\odot$ to $\approx 10^{10} \, \rm M_\odot$, with $16$ black holes exceeding masses of $10^9 \, \rm M_\odot$ \citep[e.g. see][]{Farina:22}.

Assuming the quasar is at redshift $z_{\rm QSO}$ (given in Table~\ref{tab:galaxySample}), the line-of-sight velocity shift is extracted from the galaxies' redshift $z_{\rm gal}$ by computing 
\begin{equation}
\Delta v_{\rm los} \, = \, \left(z_{\rm gal} / z_{\rm QSO} - 1\right) \left(1 + 1/z_{\rm QSO} \right)^{-1} c \, ,
\label{eq:vlos}
\end{equation}
where $c$ is the speed of light in vacuum. This equation assumes the convention that a higher-redshift galaxy companion will have a positive velocity offset \citep{Wang:23}. 
When the line-of-sight velocity offset is not directly quoted, Equation~\ref{eq:vlos} is applied to the redshift values given in the literature. This procedure converts the given redshifts into peculiar velocities assuming that the observed companion galaxy is at the same cosmological redshift as the quasar, mimicking the approach taken in observational studies \citep[e.g.][]{Decarli:17, Farina:17, Venemans:19, Wang:23}.
When the proper distances between companions and quasars are not explicitly quoted \citep[e.g.][]{Wang:23, Kashino:23}, spatial coordinates (right-ascension and declination) are converted into a proper distance $R_{\rm proj}$ as
\begin{equation}
    R_{\rm proj} \, = \, \sqrt{\left( \mathrm{RA}_{\rm gal} - \mathrm{RA}_{\rm QSO}\right)^2 \cos^2{\left( \mathrm{Dec}_{\rm gal} \right)} + \left( \mathrm{Dec}_{\rm gal} - \mathrm{Dec}_{\rm QSO} \right)^2} \, ,
    \label{eq:coordinate_conversion}
\end{equation}
where RA and Dec refer to right-ascension (in radians), respectively, and the subscripts `gal' and `QSO' refer to companion galaxy and quasar. To convert into proper distance, the same cosmological parameters as employed in the simulations (Section~\ref{sec:simulations}) are assumed.

Radial distances and line-of-sight velocities for the galaxies in the sample are shown on the left-hand panel of Figure~\ref{fig:dataset}. This presents two main features: (i) a locus of points close to $v_{\rm los} \approx 0 \, \rm km \, s^{-1}$ that broadens with decreasing radial distance and (ii) a series of data points with line-of-sight speeds that significantly deviate from this locus. Some of these systems may still not be `true' companions and instead lie several Mpc away from the quasar.

\subsection{Selecting companion galaxies}
There is some ambiguity in determining whether a source is a true galaxy companion, especially if this lies close to the quasar host galaxy \citep[e.g.][]{Neeleman:21}. The typical spatial extent of the interstellar medium of $z > 6$ quasar host galaxies is a few $\rm kpc$. If a separate source is detected at comparable scales, it may be a companion galaxy, but it may also be a substructure within the host galaxy if this is clumpy or disturbed \citep[e.g.][]{Venemans:19}. For rest-frame UV/optical emission, separate structures could also emerge as a consequence of inhomogeneous dust distribution \citep{DiMascia:21}.

Uncertainties in classifying companions exist also on the theoretical end. When does a merging satellite cease to be an independent companion to become a substructure of the massive quasar host galaxy? 
In cosmological simulations, galaxies are typically identified using \texttt{SUBFIND} \citep{Springel:01}. \texttt{SUBFIND} identifies locally overdense, gravitationally-bound structures. This algorithm may not classify galaxies which are strongly interacting with the quasar host as independent companions, treating them instead as part of the central galaxy. In \mbox{Section~\ref{sec:comparison}}, it is shown that \texttt{SUBFIND}, used in this paper to identify companions, picks up isolated systems down to radial distances of $\gtrsim 5 \, \rm kpc$. This scale roughly matches the size of the quasar host galaxies present in the simulations and defines the minimum radial threshold above which comparison between theoretical predictions and observations is less ambiguous.

\begin{figure*}
    \centering
    \includegraphics[width=0.45\textwidth]{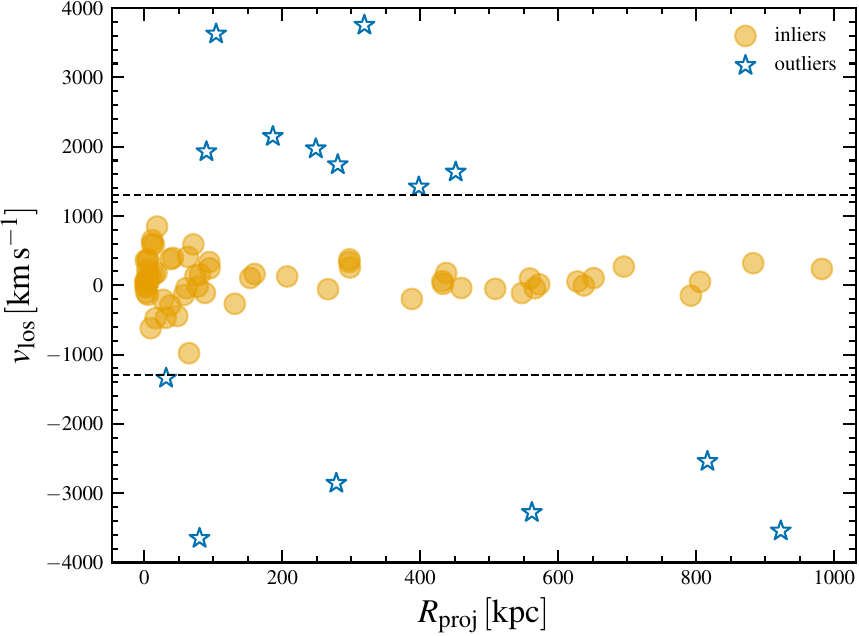}
    \includegraphics[width=0.45\textwidth]{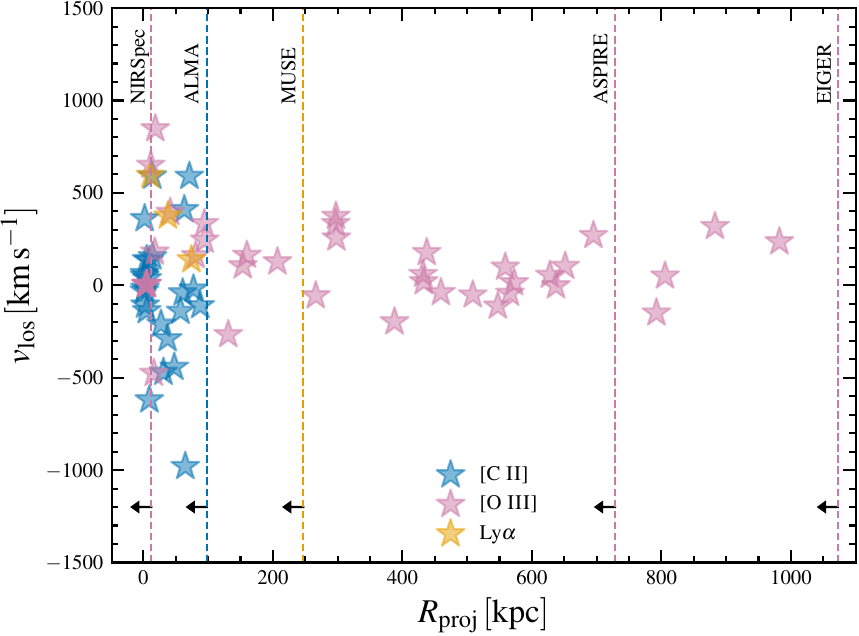}
    \caption{Line-of-sight velocity distribution of observed quasar companion galaxies as a function of projected radius. Data is taken from \citet{Decarli:17}, \citet{Farina:17}, \citet{Neeleman:19, Neeleman:21}, \citet{Venemans:19, Venemans:20}, \citet{Meyer:22}, \citet{Wang:23}, \citet{Kashino:23} and \citet{Marshall:23}. The left-hand panel separates data points according to their modified Z-score. Points with a modified Z-score $> 3$ are considered outliers (blue stars). Orange points have modified Z-scores $< 3$ and are considered inliers.  The right-hand panel zooms-in on the velocity distribution of quasar companion galaxies which are, on the basis of outlier analysis, assumed to be physically associated to the quasar host galaxy. The `Fingers of God' effect, i.e. the broadening of the line-of-sight velocity distribution with decreasing radius is particularly clear. The marker colours identify the emission line with which the companion was detected. Vertical, dashed lines give the approximate maximal radial distance between quasar and companions for different instruments.}
    \label{fig:dataset}
\end{figure*}

In order to avoid selecting ambiguous `companions', two cuts are applied to the data set presented in Section~\ref{sec:dataset}. Observational data points are considered only if the associated projected radial distances from the quasar exceed $5 \, \rm kpc$. It is worth pointing out that sources identified as companions within smaller projected radii typically have low line-of-sight speed offsets $\lesssim 350 \, \rm km \, s^{-1}$. To coalesce with the central, satellites have to dissipate their kinetic energy, which may explain why close-in systems typically have low line-of-sight velocity offsets. Such companions, however, exist in both theoretical samples (see Section~\ref{sec:results}) and cannot be used to distinguish `low-mass' and `high-mass' haloes.

Secondly, to exclude galaxies that are unlikely to be physically associated to the quasar host galaxy, observed data are only used if the line-of-sight speed offsets from the quasar is lower than $\approx 1300 \, \rm km \, s^{-1}$. There is no clear-cut threshold that distinguishes systems that have redshift offsets due to peculiar velocities from those which are simply at a different cosmological redshift. The choice of the $1300 \, \rm km \, s^{-1}$ threshold used here is grounded on three reasons:
\begin{enumerate}
    \item it is comparable to the maximum line-of-sight velocity obtained in even the theoretical `high-mass' sample (Figure~\ref{fig:vlos_rproj}). If physically-associated to the quasar, higher-velocity satellites would require even higher central halo masses than considered in this paper, placing them even more at odds with the `low-mass' sample,
    \item it is close to the threshold suggested by a computation of modified Z-scores. This is a standardised measure for outlier detection, which calculates every data point's deviation from the median, scaling it by the median absolute deviation of the whole data set. This measure is multiplied by $\approx 0.6745$ to achieve a similar scale as the standard normal distribution. The modified Z-score is computed in two different radial bins separated at $R_{\rm proj} \, = \, 100 \, \rm kpc$, since this scale approximately corresponds to the radius at which the velocity dispersion profile changes. Data points are considered outliers if the modified Z-score $> 3$. The left-hand panel of Figure~\ref{fig:dataset} shows the results of this analysis, marking inliers with yellow circles and outliers with blue stars,
    \item visually, it can be seen to separate well the locus of systems with low $v_{\rm los}$ from various isolated galaxies with much higher $v_{\rm los}$.
\end{enumerate}
The radial distance and line-of-sight velocity cuts applied here reduce the sample size to 55 galaxy companions. To further minimise uncertainties around the line-of-sight speed cut-off threshold, I explore varying this within the range $1000 \-- 2000 \, \rm km \, s^{-1}$ in my analysis.

The right-hand panel of Figure~\ref{fig:dataset} shows more clearly how inliers are distributed spatially and in velocity space. Different colours show how each galaxy companion was identified, distinguishing those detected by \cii \, emission (blue), \oiii \, emission (pink) and Ly$\alpha$ (orange).
This plot makes the `Fingers of God' effect particularly clear: the line-of-sight velocity profile is narrow at projected radii $\gtrsim 200 \, \rm kpc$ and becomes gradually broader with decreasing radius.

The vertical lines on the right-hand panel of \mbox{Figure~\ref{fig:dataset}} mark an estimate for the maximum expected radial distance between quasar host galaxy and companions for different instruments. The maximum radial distance is estimated from the FoVs of the various instruments used to identify companions, assuming the quasar is at the centre of the image.
The limited FoV of ALMA makes detection of \cii \, emitters at large radii challenging, explaining why these cluster closer to the quasar. Thanks to a larger FoV, JWST detections of \oiii \, emitters are possible out to much larger radii.

\section{Results}
\label{sec:results}
\subsection{Quasar companions }
\label{sec:bigcompanions}

\begin{figure*}
\includegraphics[width=0.95\textwidth]{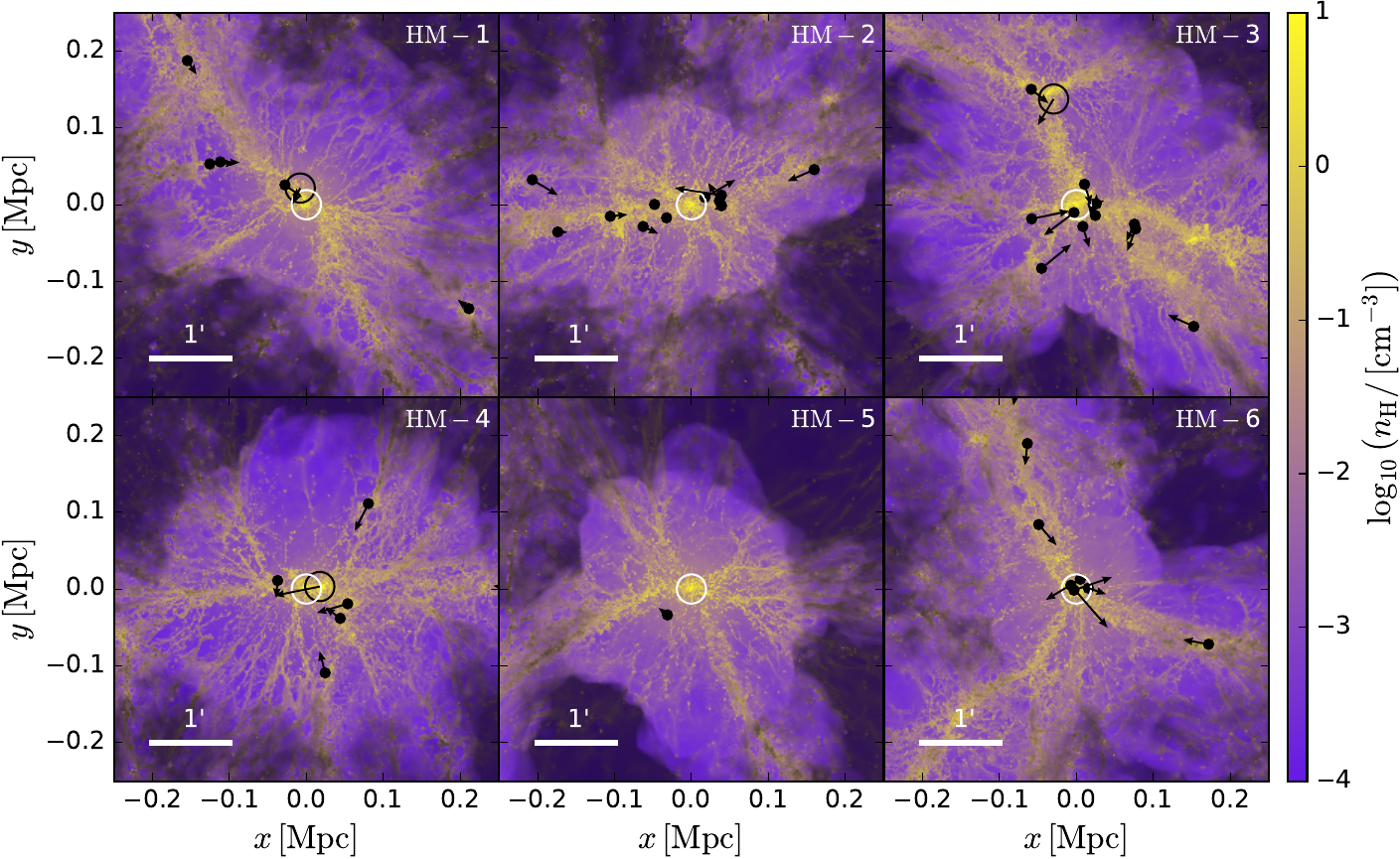}
\includegraphics[width=0.95\textwidth]{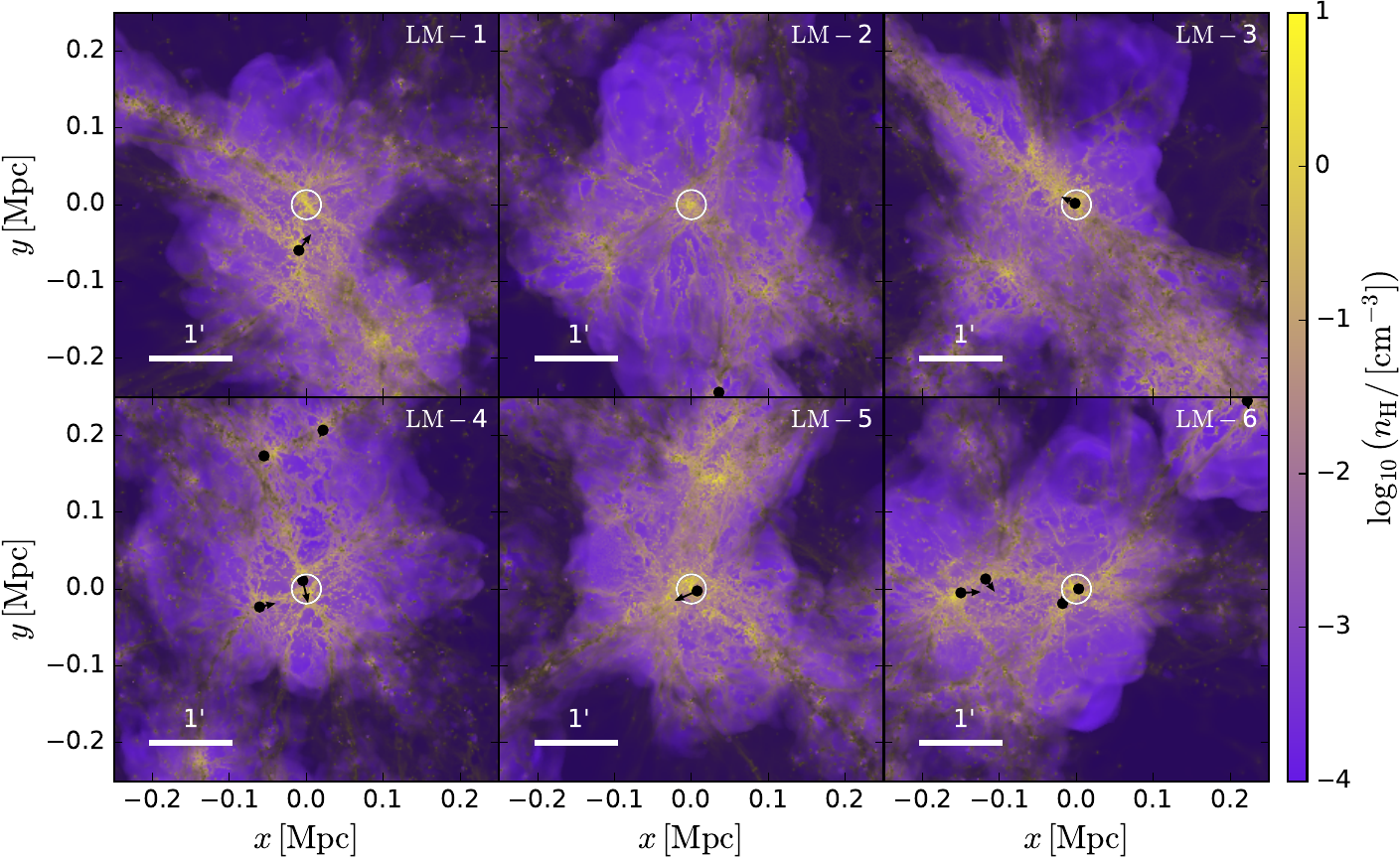}
    \caption{Density field surrounding the six haloes of the `high-mass' sample (two top rows) and the `low-mass' sample (two bottom rows) at $z \, = \, 6$. Information about ambient gas temperature is provided through the brightness of the map, with bright areas representing regions embedded within bubbles with $T \sim 10^6 \, \rm K$. Images are $0.5 \, \rm Mpc$ across. Open black circles mark the location of satellite galaxies with stellar mass $M_\star > 10^{10} \, \rm M_\odot$. Filled black circles mark the location of satellite galaxies with stellar mass $10^9 \, \mathrm{M_\odot} < M_\star < 10^{10} \, \rm M_\odot$. The galaxy hosting the most massive black hole is marked with a white circle. The arrows give the projected velocity, with their length indicating the speed. The white bar marks the spatial scale corresponding to one arcminute. The position of satellites generally traces the cosmic web and these are often, though not always, arranged into filaments. There is significant variance in the number of companions across the sample, even though all haloes within any given sample have a similar virial mass.}
    \label{fig:environment}
\end{figure*}

The large-scale distribution of gas around massive haloes at $z > 6$ has been investigated in several studies \citep{Dubois:12, DiMatteo:12, Costa:22}. Inflowing gas is concentrated along multiple, narrow streams of cool gas at $T \sim 10^4 \, \rm K$. These streams feed the central galaxy following nearly radial orbits. The typical density field is shown in Figure~\ref{fig:environment}. The two rows at the top correspond to `high-mass' halo environments, while the two bottom rows correspond to `low-mass' environments. The positions of companion galaxies are marked with filled circles (for systems with $10^9 \, \mathrm{M_\odot} < M_\star < 10^{10} \, \rm M_\odot$), open circles (galaxies with $M_{\rm \star} > 10^{10} \, \rm M_\odot$) while the galaxy hosting the most massive black hole is marked with a white circle.

In the `high-mass' sample, companion galaxies are themselves often, though not always, arranged into filaments, tracing the cool gas streams. Such `satellite filaments' are particular clear in \texttt{HM-1} and \texttt{HM-2}, where they exceed spatial scales $\sim 0.5 \, \rm Mpc$, comparable to the field of view of a single NIRCam imaging channel. This result may explain the detection of a satellite filament in the $z \, = \, 6.6$ quasar J0305, as reported by \citet{Wang:23}.
In other regions (e.g. \texttt{HM-3}), however, satellites are arranged more isotropically, so a variety of satellite configurations is expected.

Except for \texttt{HM-2}, all simulated `high-mass' quasar host galaxies possess one or more companions with $M_{\rm \star} > 10^{10} \, \rm M_\odot$ at $z \, = \, 6$ (see Table~\ref{tab:number_counts} for detailed number counts). Figure~\ref{fig:environment} shows examples of such massive companion galaxies within several $\sim 100 \, \rm kpc$ from the quasar (in addition to the quasar host galaxy).
These galaxies, however, often lie at even greater distances from the quasar host and cannot be seen in Figure~\ref{fig:environment}.
The number of such systems rises to five within $2 \, \rm Mpc$ in \texttt{HM-3} and to$10$ in \texttt{HM-6} (see Table~\ref{tab:number_counts}), pointing to significant variance in bright galaxy number counts in $z > 6$ quasar fields (see Section~\ref{sec:luminosity_functions}). 

\begin{figure*}
    \includegraphics[width=0.9\textwidth]{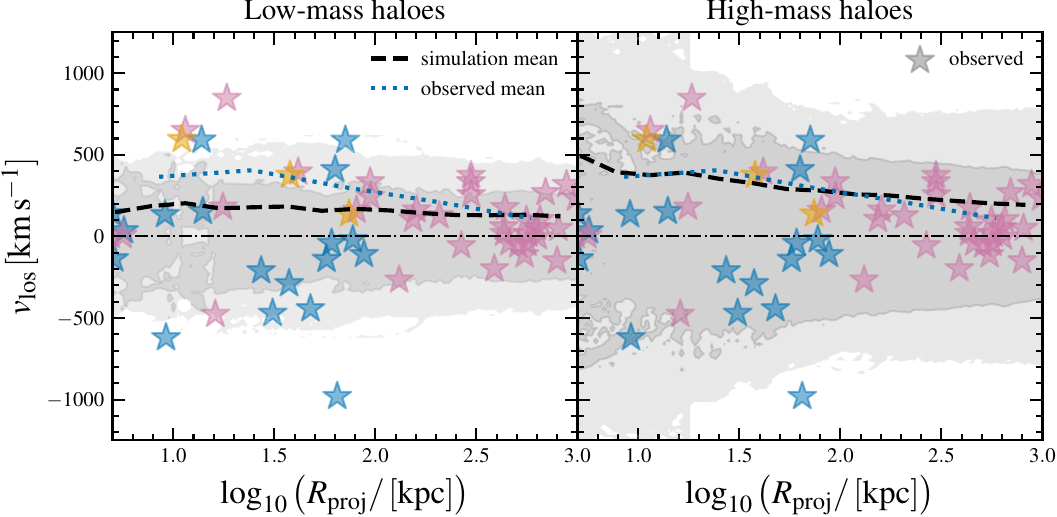}
    \caption{Theoretical distribution (shaded regions) of line-of-sight velocities for satellites with $M_{\star} > 10^{8} \, \rm M_\odot$ as a function of projected radial distance from the most massive black hole, for the `low-mass' (left-hand panel) and `high-mass' (right-hand panel) halo samples. This distribution is obtained by stacking the individual distributions found for 10,000 random lines-of-sight for each of the six target haloes in either sample at $z \, = \, 6$. The velocity distribution in every radial bin is normalised such that its integral is unity. The outer boundary of the light gray region encloses $90\%$ of the distribution and the outer boundary of the darker region encloses $50\%$ of the distribution. There is a profound difference between the line-of-sight velocity distribution of `low-mass' and `high-mass' samples. Satellite line-of-sight velocities are up to a factor two higher in the `high-mass' sample, reaching values of $1000 \-- 1250 \, \rm km \, s^{-1}$ particularly at $R_{\rm proj} \sim 10 \, \rm kpc$. Existing data (shown with stars following the colour convention of Figure~\ref{fig:dataset}) more closely follow the distribution predicted for the `high-mass' sample.}
    \label{fig:vlos_rproj}
\end{figure*}

In contrast, beyond the system hosting the most massive black hole, the entire `low-mass' sample and all its spatial volume hosts one single companion galaxy with $M_{\rm \star} > 10^{10} \, \rm M_\odot$. The discovery of bright companions with dynamical masses $> 10^{10} \, \rm M_\odot$ in the vicinity of $z > 6$ quasars \citep[e.g.][]{Decarli:17} would thus provide support to a `high-mass' host scenario. 
However, Table~\ref{tab:number_counts} also shows that the picture becomes much more ambiguous for lower-mass companions: the environments of low-mass haloes $\texttt{LM-3}$, $\texttt{LM-5}$ and $\texttt{LM-6}$ possess a comparable or greater number of neighbours with $M_{\rm \star} > 10^{9} \, \rm M_\odot$ than high-mass haloes $\texttt{HM-1}$, $\texttt{HM-3}$, $\texttt{HM-4}$ and $\texttt{HM-5}$. Haloes with $M_{\rm vir} \, \approx \, (6 \-- 7) \times 10^{12} \, \rm M_\odot$ thus often possess just as many companions as haloes with $M_{\rm vir} \, \lesssim \, 10^{12} \, \rm M_\odot$.

\subsection{Satellite kinematics}
\label{sec:kinematics}
The arrows shown in Figure~\ref{fig:environment} indicate the velocity of companion galaxies. Particularly in the `high-mass' halo environments, one can see that the velocities of galaxies located at scales $\gtrsim 100 \, \rm kpc$ mostly point towards the black hole host system. At these scales, galaxies drift in towards the quasar host halo.
As these galaxies approach the centre, they virialise: their velocities increase and become more random. A comparison with the velocity vectors shown for the `low-mass' environments further indicates that the deeper gravitational potential wells of `high-mass' haloes accelerate companions to systematically higher speeds.

A computation of the radial velocities of individual galaxies in the frame of the most massive black hole host galaxy confirms that most companions are infalling. Though they do not much exceed radial velocities\footnote{A minus sign is used to denote systems which fall in towards the galaxy hosting the most massive black hole.} of $-500 \, \rm km \, s^{-1}$ in the `low-mass' sample, infall velocities rise to $-1000 \, \rm km \, s^{-1}$ in the `high-mass' sample.
In order to understand the associated observational signature, the shaded areas of Figure~\ref{fig:vlos_rproj} give the theoretical distribution of satellite \textit{line-of-sight peculiar velocities} ($v_{\rm los}$) as a function of projected radial distance ($R_{\rm proj}$). This distribution is obtained by stacking the individual distributions of 10,000 lines-of-sight drawn in random directions for each of the six haloes in the two halo samples. Two random numbers are drawn for each line-of-sight: an angle $\phi$, sampled uniformly in the interval $[0, 2\pi]$ and $\cos{\theta}$, sampled uniformly in the interval $[-1, 1]$. These two angles are used to perform a random rotation to the coordinate system, always keeping the quasar host galaxy at the origin. 
Figure~\ref{fig:vlos_rproj} thus accounts for both (i) halo-to-halo variations and (ii) line-of-sight effects. 
As in Figure~\ref{fig:dataset}, the distribution shown in Figure~\ref{fig:vlos_rproj} excludes contributions from the Hubble flow to line-of-sight velocities. This assumption is justified because (i) observed line-of-sight velocities given in the literature are commonly derived \textit{assuming that the quasar host galaxy and companions are at the same cosmological redshift} (see Section~\ref{sec:dataset}), (ii) the Hubble flow only becomes important at large radial distances $\gtrsim 500 \, \rm kpc$, and (iii) the results of this paper depend mainly on the velocity distribution within $R_{\rm proj} \lesssim 100 \, \rm kpc$, where differences between `low-mass' and `high-mass' environments are most pronounced.
A more sophisticated approach requires the reconstruction of a light-cone, which would require a very large cosmological volume and is not viable with the small `zoom-in' simulation volumes investigated here.

Observations do not uniformly sample the radial dimension. In order to focus on the line-of-sight velocity (and not radial) distribution in the comparison between simulation and observed data, the simulation velocity distribution in Figure~\ref{fig:vlos_rproj} is normalised such that its integral in every radial bin is unity. The shaded regions thus show how line-of-sight velocities are independently distributed within any given radial bin, but contain no information about the radial distribution, which cannot be compared fairly between the  simulation and observational samples.

The typical line-of-sight velocity profile of galaxies surrounding the `low-mass' halo sample is shown on the left-hand panel of Figure~\ref{fig:vlos_rproj}. Maximum line-of-sight speeds do not typically exceed $\approx 500 \, \rm km \, s^{-1}$. In addition, the line-of-sight velocity distribution does not change significantly with radius. Instead, line-of-sight velocities are much higher for the `high-mass' sample, shown on the right-hand panel of Figure~\ref{fig:vlos_rproj}. While typical values are $\approx 300 \-- 800 \, \rm km \, s^{-1}$, satellites sometimes reach line-of-sight speeds $\gtrsim 1000 \, \rm km \,s^{-1}$. These extreme values occur mainly at $R_{\rm proj} \sim 10 \, \rm kpc$, but are possible at much larger scales $\sim 10 R_{\rm vir} \sim 1 \, \rm Mpc$ as well.

Besides the presence of extreme line-of-sight velocities, the satellite velocity distribution in the `high-mass' sample shows two prominent features: two `bands' that, starting at $R_{\rm proj} \sim 100 \, \rm kpc$, diverge towards lower radii.
The mean line-of-sight speed (dashed curve) thus rises monotonically towards the quasar host galaxy, reaching $\approx 550 \, \rm km \, s^{-1}$ in the centre. For the `low-mass' sample, mean line-of-sight speeds rise only to $\approx 200 \, \rm km \, s^{-1}$ at $\sim \, \rm kpc$ scales. The shape of the line-of-sight velocity profile reflects the two separate regimes in companion kinematics described above: (i) radial infall towards the quasar host halo at $R_{\rm proj} \gtrsim 100 \, \rm kpc$ and (ii) the virialisation of satellite galaxies and consequent rise in line-of-sight velocity dispersion at $R_{\rm proj} \lesssim R_{\rm vir} \sim 100 \, \rm kpc$.

For ease of comparison with future observational measurements, I here provide a linear fit for the mean (absolute) line-of-sight velocities in the two different halo samples, i.e. assuming that this follows the form 
\begin{equation}
|v_{\rm los}| \, = \, \alpha \log_{\rm 10}{\left( R_{\rm proj} / \mathrm{[kpc]} \right)} + \beta \, .
\end{equation}
For `low-mass' hosts, best-fit parameters are $\alpha = -47.8  \, \rm km \, s^{-1}$ and $\beta \, = \, 253.0 \, \rm km \, s^{-1}$. For `high-mass' hosts, these parameters become $\alpha = -173.3  \, \rm km \, s^{-1}$ and $\beta \, = \, 664.0  \, \rm km \, s^{-1}$. The R-squared values are, respectively, $0.79$ and $0.98$, indicating that the linear models provide good fits. Note that the fitting parameters given here are only valid in the radial interval $[5 \, \mathrm{kpc}, 1000 \, \mathrm{kpc}]$.

The analysis presented here suggests that measurements of line-of-sight velocities can be used to distinguish between host halo masses of $\lesssim 10^{12} \, \rm M_\odot$ and $\gtrsim 5 \times 10^{12} \, \rm M_\odot$ at $z \, = \, 6$. 
In the next Section, I therefore compare these predictions with existing observational data.

\subsection{Comparison with observations}
\label{sec:comparison}

Stacking all observed data points produces the  line-of-sight velocity distribution shown with blue stars in Figure~\ref{fig:vlos_rproj} (see also Figure~\ref{fig:dataset}).
These data points show striking similarities with the profile predicted for the simulated `high-mass' sample. At projected radii $> 100 \, \rm kpc$, they cluster within $300 \-- 400 \, \rm km \, s^{-1}$ from the quasar. Starting at $R_{\rm proj} \sim 100 \, \rm kpc$, the line-of-sight velocity can be seen to broaden towards lower radii (see also Figure~\ref{fig:dataset}). The mean line-of-sight velocity offset (blue dotted curve) thus increases, closely mimicking the theoretical expectation (black dashed curve). Towards the centre, observed velocity offsets reach $\approx 800 \-- 1000 \, \rm km \, s^{-1}$, showing that observed companions do sometimes reach extreme line-of-sight velocities, as expected for a `high-mass' halo. In contrast, the theoretical distribution for the `low-mass' sample reproduces neither the rise in the line-of-sight velocity dispersion nor the highest line-of-sight velocities seen in the data. 

\begin{figure*}
    \includegraphics[width=0.45\textwidth]{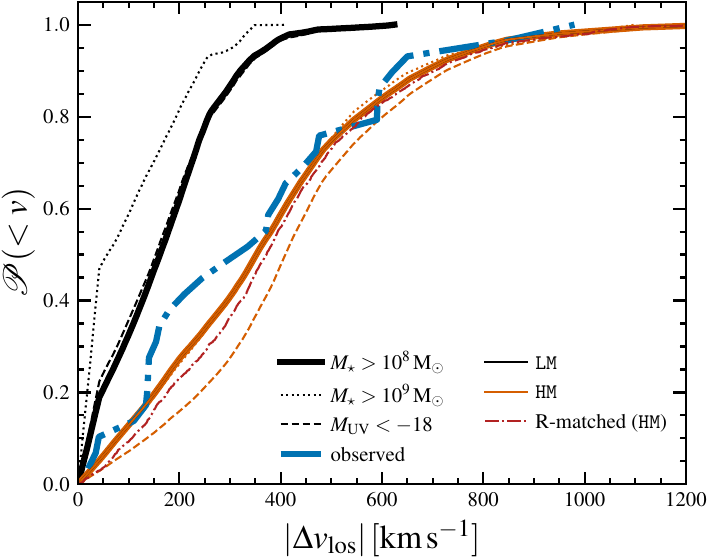}
    \includegraphics[width=0.45\textwidth]{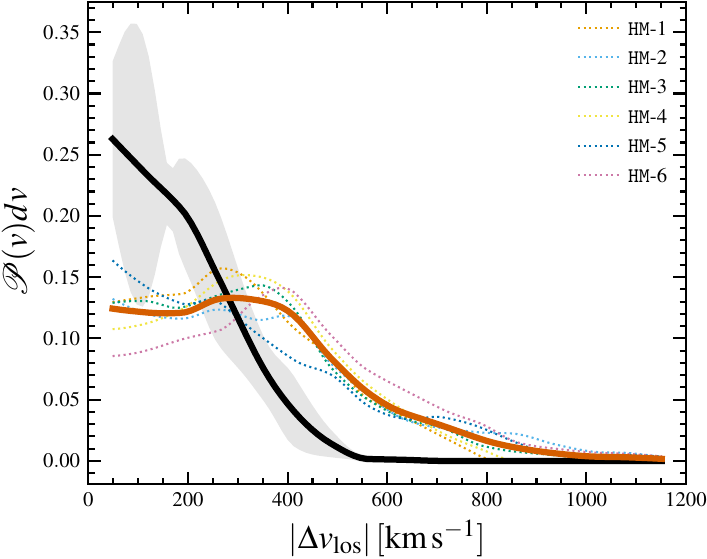}

    \caption{Left Panel: Fraction of galaxy companions within a projected radial distance $100 \, \rm kpc$ below any given line-of-sight speed. Orange curves show results for `high-mass' host haloes and black curves for `low-mass' host haloes. The red, dot-dashed curve shows the velocity distribution obtained from a radially-matched sample of galaxies around `high-mass' haloes (see text). Different line-styles reflect different companion selection criteria, such as a stellar mass selection (solid and dotted curves) and a UV luminosity selection (dashed curve). Regardless of how satellite galaxies are selected, the median line-of-sight offset from the most massive galaxy is $<200 \, \rm km \, s^{-1}$ around `low-mass' host haloes, but $\approx 350 \-- 500 \, \rm km \, s^{-1}$ in `high-mass' haloes. About one in ten satellites should have velocity offsets $> 800 \, \rm km \, s^{-1}$ in massive host haloes. Thin and thick blue, dot-dashed curves respectively show the observational distributions excluding and including Ly$\alpha$ emitters from the sample. \textit{Comparison with the theoretical distributions reveals that observational data strongly favour a `high-mass' host halo}. Right Panel: The line-of-sight speed distribution within a projected radial distance $100 \, \rm kpc$. Thick lines give the median obtained from stacking 10,000 random lines-of-sight for each halo environment from either sample. Dotted curves show median distributions for 10,000 lines-of-sight, however, from single halo environments in the \texttt{HM} sample. Shaded regions give show the $1\sigma$ scatter level. Despite significant variety in line-of-sight speed distributions, halo-to-halo variations are small when compared to the effect of switching to a lower halo mass.}
    \label{fig:probability}
\end{figure*}

A Kolmogorov-Smirnov (KS) test strengthens this conclusion. First, 1000 bootstrap samples of the same size as the observational data set are constructed by drawing points at random from the theoretical distribution (Figure~\ref{fig:vlos_rproj}) at a matching radius.
Since both the theoretical and observed distributions are symmetric around zero, the absolute values of the line-of-sight velocities are assumed. For each bootstrap sample, the KS statistic is then computed to test the null hypothesis that this sample is randomly drawn from the same underlying distribution as the observed data. The median p-value of $0.056$ found across the `low-mass' bootstrap samples all but rejects the null hypothesis. For the `high-mass' sample, a median p-value of $0.326$ is obtained -- the hypothesis that both data sets sample the same underlying distribution cannot be ruled out in this case. The p-value is sensitive to the radial and line-of-sight velocity thresholds used to select the data. Filtering out all companions with $|v_{\rm los}| > 2000 \, \rm km \, s^{-1}$ ($|v_{\rm los}| > 1000 \, \rm km \, s^{-1}$), i.e. instead of using Z-score-based selection described in Section~\ref{sec:observations}, gives a p-value of $0.513$ ($0.326$) for the `high-mass' sample and $9 \times 10^{-3}$ ($0.056$) for the `low-mass' sample. 
Sticking to the Z-score-based selection and restricting the analysis to systems within $R_{\rm proj} < 100 \, \rm kpc$ gives median p-values of $0.791$ for the `high-mass' sample and $5 \times 10^{-3}$ for the `low-mass' sample, ruling out a halo mass $\lesssim 10^{12} \, \rm M_\odot$.
If Ly$\alpha$ emitters, whose redshifts may be more uncertain and affected due to asymmetric intergalactic absorption, are excluded, the median p-value becomes $0.732$ for high-mass- and $0.018$ for low-mass samples, producing no qualitative difference in the results.
If the radial interval is narrowed to $10 \, \mathrm{kpc} < R_{\rm proj} < 100 \, \mathrm{kpc}$ (instead of using $5 \, \rm kpc$ as a lower bound), the p-value becomes $0.686$ for the high-mass sample and $0.012$ for the low-mass sample.
The conclusion that available data strongly favour a `high-mass' halo appears to be robust.

\subsection{Extreme line-of-sight velocities}
\label{sec:extremevlos}
How frequently should extreme line-of-sight offsets be detected in satellites surrounding $z > 6$ quasars? The left-hand panel of Figure~\ref{fig:probability} shows the probability of finding a system with a line-of-sight speed offset lower than some threshold value $|v_{\rm los}|$. Shown in Figure~\ref{fig:probability} is the median obtained from a stack consisting of 10,000 random lines-of-sight for each of the six haloes in either sample.
In order to emphasise differences between the theoretical distributions for the two different halo samples (see Figure~\ref{fig:vlos_rproj}), only sources within a projected radius of $100 \, \rm kpc$ from the central quasar are shown. 
Different line-styles, which correspond to different galaxy selection criteria, suggest that the probability distribution is mostly independent of how companions are selected. Two stellar mass selection cuts, namely $M_\star > 10^8 \, \rm M_\odot$ (solid curve) and $M_\star > 10^9 \, \rm M_\odot$ (dotted curve) and a UV magnitude\footnote{See Section~\ref{sec:luminosity_functions} for a description of how UV luminosities are extracted.} cut $M_{\rm UV} < -18$ (dashed curve) are explored. For `low-mass' host haloes (black curves), about $50\%$ of satellites should have line-of-sight velocities $\lesssim 100 \-- 200 \, \rm km \, s^{-1}$ and about $90 \%$ of satellites should have $|v_{\rm los}| \lesssim (300 \-- 350) \, \rm km \, s^{-1}$. Around `high-mass' host haloes (orange curves), however, $50\%$ of satellites should have line-of-sight velocities $\lesssim 450 \, \rm km \, s^{-1}$, $10 \%$ should show velocity offsets $\gtrsim 800 \, \rm km \, s^{-1}$ and a few percent should reach speeds $\approx 1000 \, \rm km \, s^{-1}$. One in ten companions should have an extreme line-of-sight velocity offset with respect to the quasar. The blue, dot-dashed curves show the distributions obtained from observed data points with $R_{\rm proj} < 100 \, \rm kpc$. \textit{A comparison between the observational and theoretical distributions shows remarkable agreement for `high-mass' host haloes}. 

For the orange curve, which gives the line-of-sight velocity distribution for companions with $R_{\rm proj} < 100 \, \rm kpc$, no attempt is made to match the projected radii of simulated and observed quasar companions. The red, dot-dashed curve on the left-hand panel of Figure~\ref{fig:probability} shows the velocity distribution obtained from a radially-matched sample of galaxies around `high-mass' haloes. This radially-matched sample is obtained as follows: for every observed companion projected radius $R_{\rm proj}$, 100 simulated companions (with $M_\star > 10^8 \, \rm M_\odot$ and within $0.5 \, \rm kpc$ from each value of $R_{\rm proj}$) are selected at random (out of the $10000$ line-of-sight realisations). The close agreement with observed data remains for this radially-matched sample.

Note that the statement that about one in ten detected satellites should exhibit an extreme line-of-sight velocity holds in a statistical sense. The right-hand panel of Figure~\ref{fig:probability} shows both the median line-of-sight speed distribution (thick curves) as well as individual distributions based on single environments from the `high-mass' sample (dotted curves). Though most distributions peak at about $\approx (300 \-- 400) \, \rm km \, s^{-1}$, there is significant halo-to-halo variation in the satellite line-of-sight velocity distribution. In the case of \texttt{HM-5}, there is even no clear peak. However, halo-to-halo variations are small when compared to the effect of switching to a lower halo mass (black curve). Above line-of-sight-velocities $\approx 400 \, \rm km \, s^{-1}$, there is no overlap between the velocity distributions for both halo samples.

\subsection{UV luminosity functions and [O{\sc III}] number counts}
\label{sec:luminosity_functions}
In order to provide an estimate of \textit{the excess} number of systems in the vicinity of bright quasars with respect to less biased regions, I here present luminosity functions for the different halo mass samples.

Far-UV luminosities are computed for every stellar particle at a wavelength of 1500 \AA \, using {\sc BPASS} \citep{Eldridge:17, Stanway:18}, a state-of-the-art Binary Population and Spectral Synthesis suite of binary stellar evolution models. BPASS is used with aid of its public Python interface \texttt{hoki} \citep{Stevance:20}. To ensure consistency with the star formation model adopted in the simulations (see Section~\ref{sec:simulations}), a Chabrier initial mass function \citep{Chabrier:03} truncated at a maximum stellar mass of $100 \, \rm M_\odot$ is assumed. The individual luminosities of each stellar particle are then added up for every subhalo, giving an estimate for the total UV luminosity for every galaxy. Dust absorption, thought to play a particularly important role in suppressing the luminosity function at $M_{\rm UV} \lesssim -21$ \citep[e.g.][]{Bouwens:21}, is modelled via an empirical relation between dust attenuation and stellar mass at $z \, = \, 6$ \citep{Bogdanoska:20}. This approach circumvents the need for additional assumptions (e.g. dust-to-gas-ratio, dust opacity) required to directly extract dust extinctions from the simulations. 
To compute total \oiii\, luminosities (i.e. at both 4960 \AA\, and 5008 \AA), the UV-\oiii \, conversion fit of \citet{Matthee:23}, given as
\begin{equation}
    \log_{\rm 10}{L_{\rm [OIII]} / \mathrm{[erg \, s^{-1}]}} \, = \, 42.60 - 0.3 (M_{\rm UV} + 20)\, ,
    \label{eq:oiiiconversion}
\end{equation}
is assumed together with a scatter level of $0.27 \, \rm dex$ \citep{Matthee:23}. 
In Equation~\ref{eq:oiiiconversion}, $M_{\rm UV}$ is taken as the dust-attenuated luminosity at 1500 \AA \, computed above. To extract \oiii \, luminosities at 5008 \AA \, specifically, the expected intrinsic intensity ratio of 2.98 between 5008 \AA \, and  4960 \AA\, is assumed.

\begin{figure}
\includegraphics[width=0.45\textwidth]{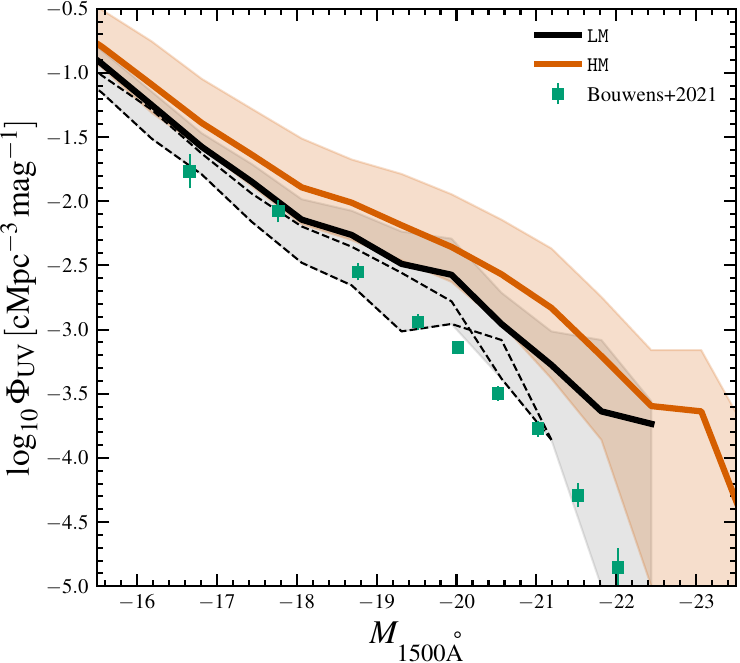}
\includegraphics[width=0.45\textwidth]{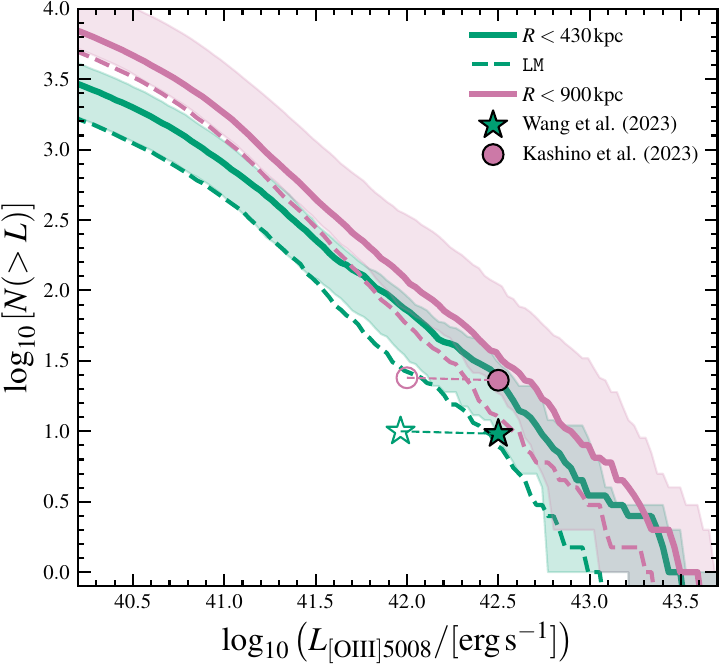}
    \caption{Top: median luminosity functions at 1500 \AA \, in `low-mass' and `high-mass' samples (thick black and orange curves, respectively) at $z \, = \, 6$. Both samples represent biased regions and thus exceed the luminosity function of \citet{Bouwens:21} (green data points), except for the two least biased volumes in the \texttt{LM} sample (grey, dashed curves). \textit{The scatter within each halo (shaded regions) sample is larger than the difference in the luminosity functions found by changing the halo mass.} Bottom: cumulative number counts of \oiii \, emitters as a function of luminosity. Different colours give expected number counts within cylindrical regions of different radii. The green curve gives the number counts in a region of comparable size as probed by \citet{Wang:23}, while the pink curve probes a larger region such as observed by \citet{Kashino:23}. Observational constraints fall within the scatter of the predicted number count for the `high-mass' sample if only the brightest sources are considered (filled markers). However, they are also compatible with a `low-mass' host halo (dashed curves). Including all detected sources (empty markers) gives number counts lower than predicted. However, observations underestimate number counts at these luminosities (see text).} 
    \label{fig:lumfunction}
\end{figure}

Luminosity functions are then computed by selecting all sources within a spherical region of radius $2 \, \rm Mpc$ centred on the most massive black hole.
The top panel of Figure~\ref{fig:lumfunction} shows the dust-attenuated UV luminosity functions at $z \, = \, 6$ obtained by taking the mean over the `zoom-in' regions over each halo mass sample. Shaded regions mark the interval between the minimum and maximum luminosity functions found across a given sample. 

The more pronounced overdensity in the `high-mass' halo sample (thick orange curve) is evident through a systematically higher normalisation than found for the `low-mass' sample (thick black curve). The offset in the normalisation between both samples grows with increasing (more negative) UV magnitude, reaching about $0.5 \, \rm dex$ at $M_{\rm UV} \, = \, -21$. At the bright end, variance also becomes highest, exceeding $\sim 1 \, \rm dex$. At every luminosity, the scatter is overwhelms the difference caused by changing the central halo mass. 

Observationally-derived luminosity functions at $z \, = \, 6$ \citep[e.g.][]{Bouwens:21} have a lower normalisation (green data points) than even the luminosity function obtained for the `low-mass' sample. This behaviour is expected because though less extreme than in the `high-mass' sample, `low-mass' haloes are still very massive in the context of the $z \, = \, 6$ Universe and their environment should still be regarded as biased. Selecting the two least massive haloes from within the `low-mass' sample (\texttt{LM-1} and \texttt{LM-2}) yields the luminosity functions shown with black, dashed curves in the top panel of Figure~\ref{fig:lumfunction}. These luminosity functions are in much closer agreement with data from \citet{Bouwens:21}.

The bottom panel of Figure~\ref{fig:lumfunction} gives the predicted median cumulative number of \oiii \, emitters as a function of \oiii \, (5008 \AA) luminosity. As explained above, this is obtained from the rest-frame UV luminosity computed in post-processing from the simulations using Equation~\ref{eq:oiiiconversion}. Source number counts exclude the quasar host galaxy and are computed within a cylinder with length $2 \, \rm Mpc$ and two different radii (different colours). 
The NIRCam observations of \citet{Wang:23} and \citet{Kashino:23} detect sources within fields of view of, respectively, $5.5 \, \rm arcmin^2$, and a rectangle with dimensions $6.5 \times 3.4 \, \rm arcmin^2$. To compare, I choose cylinder radii such that their projected areas match the area probed by \citet{Wang:23} (green), and \citet{Kashino:23} (pink). Solid curves give results for the `high-mass' sample and dashed curves for the `low-mass' sample.

Within a circular aperture with radius $430 \, \rm kpc$ (green curve), the predicted median number of companions with $L_{\rm [O {\sc III}]} > 10^{42} \, \rm erg \, s^{-1}$ is $\approx 80$.  Within a radius $900 \, \rm kpc$, this number rises to $\approx 125$.
In their NIRCam observations, \citet{Wang:23} present evidence for 10 companions down to $L_{\rm [O {\sc III}]} \approx 10^{42} \, \rm erg \, s^{-1}$ (green, empty star in Figure~\ref{fig:lumfunction}).
In their {\sc Eiger} survey, \citet{Kashino:23} find evidence for 24 companions (pink, empty circle) around J0100 at $z \, = \, 6.3$.
At first glance, these numbers appear too low when compared to predicted numbers even for the `low-mass' sample. However, \citet{Matthee:23} explain that at UV magnitudes $\gtrsim -20$ (corresponding to $\log_{\rm 10}{\left( L_{\rm [O {\sc III}]} / \mathrm{[erg \, s^{-1}]} \right)} \approx 42.6$), the {\sc Eiger} survey only selects the upper end of the equivalent width distribution. As a consequence, \citet{Matthee:23} find an order of magnitude higher number density in the UV than in \oiii \, at $M_{\rm UV} \approx -18.5$. In addition, \citet{Matthee:23}, for instance, estimate a median completeness of $79\%$ for the entire sample, though this drops to only 
$\approx 40 \%$ for \oiii \, emitters detected at $L_{\rm [O {\sc III}]} \, \approx \, 10^{42} \, \rm erg \, s^{-1}$.
The open symbols in the bottom panel of Figure~\ref{fig:lumfunction} thus probably underestimate the true number of \oiii \, emitters in the targeted quasar fields.

A more reliable number count can be estimated by focusing only on the number of systems with  $\log_{\rm 10}{L_{\rm [O {\sc III}]} / \mathrm{[erg \, s^{-1}]}} > 42.5$. Out of the 10 companions identified in \citet{Wang:23}, eight have $\log_{\rm 10}{L_{\rm [O {\sc III}]} / \mathrm{[erg \, s^{-1}]}} > 42.5$ (within uncertainty). \citet{Kashino:23} do not tabulate [O{\sc III}] luminosities and I thus assume that, like in \citet{Wang:23}, $80\%$ of the identified companions exceed the same luminosity threshold. Further assuming an incompleteness level of $20 \%$ gives the filled symbols in the bottom panel of Figure~\ref{fig:lumfunction}. Number counts can now be reconciled with theoretical values. \textit{Remarkably, however, the theoretical scatter is so large that data can hardly distinguish between `high-mass' or a `low-mass' host.}

However, the case in favour of a `low-mass' host is not strong. \citet{Wang:23} report line-of-sight velocities $\approx 670 \, \rm km \, s^{-1}$ and $\approx 890 \, \rm km \, s^{-1}$ in two companions at radial distances of $10 \-- 20 \, \rm kpc$. These values cannot be reproduced by haloes from the `low-mass' sample (see Section~\ref{sec:comparison}). 
In addition, \citet{Wang:23} point out that the same quasar where 10 \oiii \,-emitting companions are found also hosts three systems detected by ALMA through \cii \, emission. These sources are not detected by JWST due to their faint intrinsic luminosities as well as due to point-spread function (PSF) effects. Careful PSF subtraction may bring theoretical and observed numbers into closer agreement.
Other discrepancies between observations and theoretical predictions may easily be due to inaccurate modelling of feedback processes. \citet{Costa:14} show that the strength of supernova feedback can cause order of magnitude variations in quasar companion counts. Since supernova winds are not calibrated\footnote{An alternative strategy would be to calibrate the parameters of the supernova model to ensure the desired UV luminosity function is reproduced. This approach, however, comes with the risk that this is reproduced correctly for the wrong reasons, via e.g. strong supernova-driven outflows, as opposed to missing physics in the simulations.} in the simulations used here, perfect agreement with observations should not be expected.
In addition, the simulations assume a spatially uniform UV background. An inhomogeneous background, which should be enhanced around quasars, may more strongly suppress star formation in galaxies in quasar fields, though this effect is expected to be more significant in faint galaxies with $M_{\rm UV} \gtrsim -16$ \citep[e.g.][]{Chen:20}.

These considerations aside, the main point of this section is not to show that cosmological simulations are able to reproduce high-redshift luminosity functions. The aim is to show that galaxy number counts do not strongly constrain the halo mass with as much precision as the dynamics of quasar companions, as presented in Section~\ref{sec:comparison}. Companion dynamics provides a more direct probe of the quasar host galaxy potential wells and, I argue, constrain their mass more effectively than number counts. Cosmic variance often ensures that highly massive haloes disguise as lower mass haloes, a conclusion now made in a variety of studies \citep[e.g.][]{Overzier:09, Costa:14, Habouzit:19, Buchner:19, Ren:21}. Further complications, such as line-of-sight variations, add even more scatter to the observed number of companions \citep{Zana:23}.

\section{The host haloes of $z > 6$ quasars}
The analysis presented in this paper implies that a low-mass ($M_{\rm vir} \lesssim 10^{12} \, \rm M_\odot$) host scenario is not typical, as this cannot account for the observed line-of-sight velocities of quasar companion galaxies. 
A close match between the theoretical and observed line-of-sight velocity distributions is found for high halo masses of $\approx (6 \-- 7) \times 10^{12} \, \rm M_\odot$. While these masses appear to be at odds with a `low-mass' scenario, this paper's results also suggest that yet more extreme dark matter haloes, e.g. with mass $\gtrsim 10^{13} \, \rm M_\odot$, should not be the typical $z > 6$ quasar hosts. Such haloes would likely accelerate companion galaxies to even greater speeds than observed or predicted for the `high-mass' sample.

This section addresses the wider implications of these findings for the nature of black hole growth at high redshift. As a starting point, I discuss how the halo mass constraint obtained in this paper compares with the values in models that produce black holes with masses $\sim 10^9 \, \rm M_\odot$ at $z \, = \, 6$. I then discuss how this constraint compares with other observational estimates. Finally, I briefly explore the need for additional physics in models of high-redshift black hole growth.

\subsection{The need for a high halo mass in theoretical models}
Cosmological hydrodynamic simulations able to explain black hole masses $\sim 10^9 \, \rm M_\odot$ at $z \, = \, 6$ generally invoke high halo masses $\gtrsim 10^{12} \, \rm M_\odot$. \citet{Sijacki:09} and \citet{Costa:14} invoke haloes with mass $\gtrsim 5 \times 10^{12} \, \rm M_\odot$, \mbox{\citet{Zhu:22}} target a halo with mass $\sim 10^{13} \, \rm M_\odot$, while \citet{Smidt:18} and \citet{Lupi:19} simulate haloes that exceed $10^{12} \, \rm M_\odot$ already at $z \, = \, 7$.
Using the {\sc Bluetides} suite of cosmological simulations,  \citet{Tenneti:18} consider black holes with mass $\lesssim 10^8 \, \rm M_\odot$ (lighter than considered here) at a higher redshift ($z \, = \, 8$), showing that their \textit{descendants} often do not go on to reside in the centre of the most massive haloes at $z \, = \, 0$, a point also made in \citet{Angulo:12}. However, in their Figure~4, \citet{Tenneti:18} show that the most massive black holes at $z \, = \, 8$ are hosted by the most massive haloes at that redshift. 
Using constrained Gaussian realisations to reconstruct the initial conditions of {\sc Bluetides}, \citet{Huang:20} show the same set of physics modules is able to produce black hole masses $\sim 10^9 \, \rm M_\odot$ at $z \, = \, 6$. At that time, the host halo masses are $\approx 3 \times 10^{12} \, \rm M_\odot$ (see their Figure~9), within a factor of 2 of the `high-mass' halo bin considered in this paper. 

Using the semi-analytic model {\sc Galform}, \citet{Fanidakis:13}, however, conclude that the \textit{brightest quasars} at $z \, = \, 6$ reside in haloes with masses $10^{11} \-- 10^{12} \, \rm M_\odot$, corresponding to the `low-mass' sample considered in this paper.
Note that this model still predicts that the most massive black holes reside in haloes with mass $> 10^{12} \, \rm M_\odot$ \citep[see Figure 5 in][]{Fanidakis:12}, as in this paper's `high-mass' sample. The difference is that such black holes no longer accrete efficiently and do not produce bright quasars in {\sc Galform}. This behaviour is driven by the assumption that haloes with mass $\gtrsim 10^{12} \, \rm M_\odot$ at $z \, = \, 6$ are mainly composed of hot gas with a long cooling timescale and are thus unable to feed the central galaxy and its black hole at a high enough rate to power a quasar. 
Cosmological simulations instead predict a co-existence between a hot, diffuse atmosphere and streams of cool gas even in `high-mass' haloes (Figure~\ref{fig:environment}). These streams feed the central galaxy and trigger quasar episodes. 

In summary, the new halo mass constraint obtained in this paper is broadly consistent with the typical halo masses invoked in cosmological models to successfully grow black holes to $\sim 10^9 \, \rm M_\odot$ by $z \, = \, 6$.

\subsection{Observational evidence for massive host haloes}
\label{sec:otherevidence}

It is interesting to note the agreement between the halo mass $\approx 5 \times 10^{12} \, \rm M_\odot$ suggested in this paper with the estimate of \citet{Arita:23}, where a computation of the auto-correlation function of 107 $z > 6$ quasars yields a dark matter halo mass of $\approx 5 \times 10^{12} \, h^{-1} \, \rm M_\odot \approx 6.8 \times 10^{12}\, \rm M_\odot$. Through a unified model of the quasar-galaxy cross-correlation-, galaxy auto-correlation-, quasar luminosity- and galaxy luminosity functions, \mbox{\citet{Pizzati:24}} obtain a broad host halo mass distribution ranging from $\approx 2 \times 10^{12} \, \rm M_\odot$ to $\approx 6 \times 10^{12} \, \rm M_\odot$. This halo mass range sits between this paper's `high-mass'- and `low-mass' samples, though its higher end is consistent with the halo mass $\approx 5 \times 10^{12} \, \rm M_\odot$ estimated here.

Additional evidence is provided by considering the properties of gas reservoirs around $z > 6$ quasars.
Very massive haloes at $z > 6$ can contain large cool gas reservoirs and are not necessarily composed predominantly of hot gas.
Besides hosting rapidly-accreting black holes, quasar host galaxies are typically highly-star forming \citep[e.g.][]{Bertoldi:03, Tripodi:23c}, suggesting strong cooling.
The recent detection of Ly$\alpha$ nebulae \citep{Farina:17, Farina:19, Drake:19} unambiguously shows that $z > 6$ quasars are indeed embedded within abundant cool gas reservoirs, extending out to $\gtrsim 30 \, \rm kpc$ (almost half the virial radius). \citet{Farina:19} estimate these to be sufficiently massive to feed the central galaxy at rates $\lesssim 100 \, \rm M_\odot \, yr^{-1}$. The morphology, luminosity and inferred surface brightness profiles of such Ly$\alpha$ nebulae can all be reproduced by cosmological simulations targeting $\approx 5 \times 10^{12} \, \rm M_\odot$ haloes tracing large-scale overdensities at $z \, = \, 6$ \citep{Costa:22}. The agreement between predicted and observed circum-galactic medium properties strengthens the case that $z > 6$ quasars do reside in massive haloes. 

Interestingly, the properties of observed Ly$\alpha$ nebulae can only be reproduced if AGN feedback is modelled in the cosmological simulations \citep{Costa:22}.
This ensures that neutral hydrogen and dust in the galactic nucleus are cleared out, allowing the Ly$\alpha$ flux generated in the centre to escape. AGN-powered outflows then propagate out into the halo, partially disrupting the large-scale cold flow network \citep{Dubois:13} and pushing out diffuse halo gas \citep{Costa:14b}.
Energy deposition by AGN regulates black hole growth \citep[see][for early examples]{Sijacki:09, DiMatteo:12}, but instead of a long-term halt to black hole accretion, feedback leads to a variable light curve flickering on $\lesssim \, \rm Myr$ timescales, alternating between periods of low- and high accretion rates that temporarily elevate the quasar luminosity up to values $L_{\rm QSO} \approx 10^{47} \-- 10^{48} \, \rm erg \, s^{-1}$. 

While this behaviour is plausible, there are reasons to question the AGN feedback models adopted in hydrodynamic simulations. An indication that some of these models may be insufficient is illustrated in Figure 5 of \citet[][]{Bennett:23}, which shows that about $10\% \-- 20\%$ of baryons are converted into stars in their simulations. Albeit highly uncertain at $z \, \approx \, 6$, abundance matching constraints \citep[e.g.][]{Moster:18, Behroozi:18} typically predict up to factor $\approx 5$ times lower star formation efficiencies at matching halo masses (with large scatter). \citet{Bennett:23} find host stellar masses $\approx 1 \-- 3 \times 10^{11} \, \rm M_\odot$, in agreement with e.g. \citet{vanderVlugt:19}, \citet{Lupi:19}, \citet{Huang:20}, \citet{Zhu:22} and \citet{Bhowmick:22}.  
In some simulations \citep{Barai:18, Valentini:21}, the host stellar mass is lower ($\approx 5 \times 10^{10} \, \rm M_\odot$), but this result could be explained by the choice of $\approx 5$ times lower mass haloes. 

Observational constraints of dynamical masses show small, but systematic, tension with these predictions.
\citet{Neeleman:21} model the kinematics of $z > 6$ quasar host galaxies, deriving a mean dynamical mass for quasar host galaxies of $\approx 5 \times 10^{10} \, \rm M_\odot$. Using an ALMA sample of 27 quasars, \citet{Decarli:18} estimate host dynamical masses in the range $2\times 10^{10} \-- 2 \times 10^{11} \, \rm M_\odot$. Through an analysis of the rotation curve measured for a quasar host galaxy, \citet{Tripodi:23} derive a host mass $\approx 5 \times 10^{10} \, \rm M_\odot$. There is clearly considerable spread in estimated dynamical masses and uncertainty is significant. Nevertheless, simulations appear to systematically predict quasar host galaxies with stellar masses close to the upper bracket of the observed distribution \citep[as already noticed in][]{Valiante:14}.
Using synthetic images generated from a cosmological simulation of a $z \, = \, 7$ quasar, \citet{Lupi:19} show that dynamical masses obtained from \cii \, velocity maps may underestimate the actual mass by a factor $\gtrsim 3$.
Observational bias is therefore a very plausible explanation for a discrepancy with theoretical predictions.

\subsection{Super-Eddington accretion?}
It is nevertheless interesting to explore theoretical routes to reducing the stellar masses of $z \, = \, 6$ quasar host galaxies.
For theoretical models, the challenge is to grow a black hole to $\sim 10^9 \, \rm M_\odot$, while preventing too much gas from forming stars on its descent towards the galactic nucleus. While suppressing star formation would reconcile simulated stellar masses with abundance matching constraints, explaining lower-than-expected dynamical masses (assuming these reflect the real mass of the system) would require more efficient gas ejection or the prevention of gas infall.
The solution to this problem is not trivial. Since AGN feedback affects both star formation and black hole growth, raising the AGN feedback efficiency comes at the cost of a lower self-regulated black hole mass \citep{DiMatteo:05, Springel:05b} and may prevent this to reach $\sim 10^9 \rm M_\odot$ by $z \, = \, 6$. 

\citet{Bennett:23} find that earlier onset of AGN feedback can reduce the stellar mass at $z \approx 6$ without reducing the black hole mass. This result sketches a possible solution: AGN feedback must be stronger in faint AGN in the $z \gtrsim 7$ progenitors of $z \approx 6$ quasars. Interestingly, in \citet{Bennett:23}, earlier feedback is made possible by removing the artificial limit that caps the black hole accretion rate to the Eddington limit. At $z \gtrsim 6$, this `Eddington ceiling' prevents gas from being accreted even if this is available. Instead, this gas piles-up around the black hole, where it forms stars. If the `Eddington ceiling' is lifted, the black hole can swallow some of this gas and expel the remainder as a consequence of higher energy deposition rates, lowering the stellar mass.

\citet{Massonneau:23} model super-Eddington AGN feedback in simulations of isolated disc galaxies, finding this can be so effective as to interrupt accretion at even lower black hole masses. Some of the parameter space (e.g. low black hole spins), however, appears to allow for significant black hole growth. 
It is important to point out that capturing a cosmological environment would place greater demands on the strength of AGN feedback than isolated halo conditions \citep[e.g.][]{Costa:14b}.
Thus, if applied to a cosmological setting, there is a good chance that more vigorous inflows are able to reignite black hole growth even after a powerful super-Eddington episode and that the cumulative feedback effect of multiple such outbursts is the long term suppression of cosmological inflow and stellar mass. Exploring the impact of super-Eddington feedback in cosmological simulations of $z > 6$ quasars is thus an important future research direction. 

\subsection{Outlook}
If the host halo mass of $z > 6$ quasars is settled, then what questions remain open? A strong assumption made in cosmological simulations is that the initial seed black holes are massive and already $\sim 10^5 \, \rm M_\odot$ at $z \approx 15 \-- 20$. This mass is in line with the `direct collapse' black hole seed formation channel explored in various studies \citep{Bromm:03, Volonteri:05, Regan:09, Chon:20, Latif:22}. Adopting this formation channel is, in part, driven by numerical limitations; resolving `mini-haloes' with mass $\sim 10^5 \, M_\odot$, as expected to host Pop-III stars, and their growth to a mass $\sim 5 \times 10^{12} \, \rm M_\odot$ by $z \approx 6$ would require $\sim 10^{10}$ dark matter particles \citep[comparable to the entire Illustris-TNG 300 simulation, ][]{Springel:18}. \citet{Chon:21} describe a `hostile' environment for black hole growth even if seed black holes are massive. Growth has to proceed despite supernova feedback, AGN feedback, despite the seed black hole likely forming in the more pristine outskirts of high-redshift galaxies and consequently wandering around the potential minimum where accretion would be most efficient. To circumvent such difficulties, current cosmological simulations \citep[e.g.]{Sijacki:09, Costa:14, Feng:16, Barai:18} apply a strategy whereby black hole particles are frequently repositioned at the potential minimum.

In a `survival of the fittest' scenario \citep[see also][]{Volonteri:06}, it is possible that only a fraction of the haloes targeted in this study would hold on to their seed black holes efficiently enough to ensure they grow to $\sim 10^9 \, \rm M_\odot$ by $z \, = \, 6$. 
An important future task will involve examining the $z \approx 15 \-- 30$ progenitors of the `high-mass' haloes considered in this paper and assess their likelihood to (i) form massive seed black holes either through direct collapse or through an intermediate compact nuclear cluster \citep[e.g.][]{Katz:15}, and (ii) to grow via gas accretion and black hole mergers.
In the process we will, however, be inevitably confronted with multiple thorny questions:
\begin{enumerate}
    \item How do the lighter black hole progenitors sink and stay pinned to the halo centre where they can grow efficiently?,
    \item What is the role of super-Eddington accretion and how should this be modelled?,
    \item What processes ensure that gas is funneled from pc scales down to the black hole's sphere of influence? 
\end{enumerate}

The population of faint AGN detected at $z \, = \, 4 \-- 7$ by e.g. \citet{Harikane:23} and \citet{Matthee:24} might correspond to those seed black holes that never `made it'.
Recent discoveries of black holes with masses $> 10^6 \, \rm M_\odot$ at $z \approx 10$ \citep{Maiolino:23, Bogdan:24} start probing the earliest stages of the black hole growth. If these massive black holes turn out to be numerous already at that time, both their masses and large number will pose new challenges to theoretical models.

\section{Conclusions}
This paper presents theoretical predictions for the line-of-sight velocity distribution of galaxy companions of quasars at $z > 6$. Predictions are provided for two different dark matter halo mass samples: `low-mass' host haloes with virial mass $\approx 5 \times 10^{11} \-- 10^{12} \, \rm M_\odot$ and `high-mass' haloes with mass $\approx (6 \-- 7) \times 10^{12} \, \rm M_\odot$. 
Key conclusions of this study include:
\begin{enumerate}
    \item If hosted by `high-mass' haloes, most quasar host galaxies, though not all, should possess one or multiple companions with stellar mass $> 10^{10} \, \rm M_\odot$ within a radial distance of $2 \, \rm Mpc$. Even within a halo mass-matched sample, the number of such companions can vary widely, ranging from null to $10$ (Section~\ref{sec:bigcompanions}), as a consequence of strong cosmic variance effects,
    \item The main observational signature of a `massive host' scenario consists of a line-of-sight velocity distribution that broadens with decreasing projected radius. Lower-mass host haloes produce much flatter line-of-sight velocity profiles and velocity offsets that do not exceed $\approx 500 \, \rm km \, s^{-1}$ (Figure~\ref{fig:vlos_rproj}). Existing observational line-of-sight velocities measurements from ALMA, MUSE and JWST observations show that quasar companions at $z \, = \, 6$ follow a broadening velocity profile and reach line-of-sight speeds $\sim 1000 \, \rm km \, s^{-1}$, as predicted for a `high-mass' scenario. I argue that the existence of such high line-of-sight speeds and the shape of the line-of-sight velocity profile is inonsistent with a `low-mass' scenario (Section~\ref{sec:comparison}).
    \item The observed line-of-sight velocity distribution, particularly for galaxies within projected radial distances of $\sim 100 \, \rm kpc$ of the central quasar, constrain the central halo mass more sensitively than galaxy number counts. In `high-mass' haloes, the line-of-sight velocity distribution is characterised by a peak at $\approx (350 \-- 400) \, \rm km \, s^{-1}$ and a tail extending to extreme line-of-sight speeds $\sim 1000 \, \rm km \,s^{-1}$. About $10\%$ of satellites are predicted to show velocity offsets greater than $800 \, \rm km \, s^{-1}$ (Section~\ref{sec:extremevlos}),
    \item While the UV luminosity function and the number of [O{\sc III}] emitters are, on average, enhanced in the `high-mass' halo sample, the scatter introduced by cosmic variance overwhelms the variance associated to changing the halo mass. A strong constraint on halo mass cannot be obtained from single-object observations and a stack over a large number of quasar fields would be required (Section~\ref{sec:luminosity_functions}),
\end{enumerate}

I conclude that the halo masses of quasars at $z \, = \, 6$ must be $\gtrsim 5 \times 10^{12} \, \rm M_\odot$. Observed companion galaxies are too fast for them to be explained by lower-mass haloes.
\citet{Costa:19} give a description of what is likely to happen to these companions. Satellites eventually meet their end as they approach the quasar host galaxy. Signs of tidal interactions should become evident already at scales $\approx (20 \-- 30) \, \rm kpc$. As the satellites near the centre, they become strongly tidally disrupted, leaving behind large streams of stellar and gaseous debris and, for the most tightly-bound satellites, a stellar core orbiting through a stellar halo built up via past encounters. 
Additional data will be delivered by JWST in the coming months, e.g. from campaigns such as {\sc ASPIRE} \citep{Wang:23, Yang:23} and {\sc EIGER} \citep{Kashino:23, Matthee:23, Eilers:23}. This larger statistical sample will be able to test this paper's theoretical predictions more rigorously and place even tighter constraints on the cosmic sites of $z > 6$ quasars.
  
\section*{Acknowledgements}
I gratefully thank the anonymous referee for a thoughtful and helpful report. I thank Manuela Bischetti, Roberto Decarli and Hannah Stacey for kindly assisting a helpless theorist with collecting relevant data from the observational literature. I further thank Chervin Laporte and Feige Wang for inspiring discussions that have shaped this work. Finally, I thank Martin Haehnelt, Debora Sijacki and Volker Springel for helpful comments on the manuscript.

\section*{Data Availability}
 
Data will be shared upon reasonable request.



\bibliographystyle{mnras}
\bibliography{lit} 

\bsp	
\label{lastpage}
\end{document}